\documentclass[amssymb,aps,prc,nofootinbib,superscriptaddress]{revtex4-2}
\usepackage{graphicx}
\usepackage{subcaption} 
\usepackage{mathtools}
\usepackage{physics}
\usepackage{slashed}
\usepackage{url}
\usepackage[normalem]{ulem}
\usepackage{caption}
\usepackage{enumitem}
\usepackage[colorlinks=true, linkcolor=blue, citecolor=red, urlcolor=red]{hyperref}    
\usepackage[percent]{overpic}
\usepackage{float}
\newcommand{\be}{\begin{eqnarray}}
	\newcommand{\ee}{\end{eqnarray}}

\newcommand{\bfz}{{\bf 0}_{\perp}}

\newcommand{\bfk}{{\bf k}_{\perp}}

\newcommand{\bfkj}{{\bf k}_{\perp j}}

\newcommand{\bfP}{{\bf P}_{\perp}}

\usepackage{xcolor}
\newcommand{\orcid}[1]{\href{https://orcid.org/#1}{\includegraphics[width=8pt]
{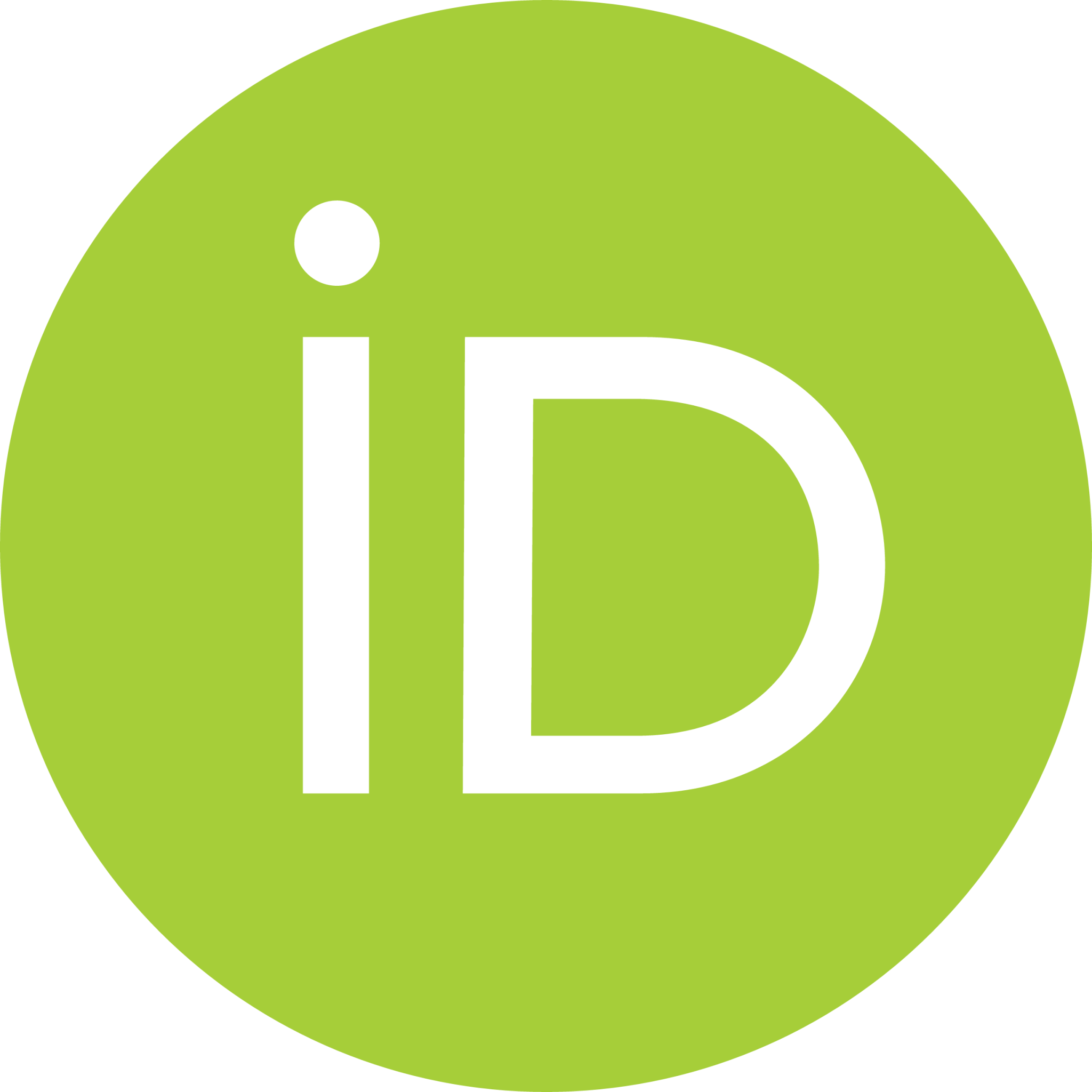}}}
\begin{document}
\title{Valence quark distribution of the pion inside a medium with finite baryon density: A Nambu--Jona-Lasinio model approach}

\author{Ashutosh Dwibedi\orcid{https://orcid.org/0009-0004-1568-2806}}
\email{ashutoshdwibedi92@gmail.com}	
\affiliation{Department of Physics, Indian Institute of Technology Bhilai, Kutelabhata, Durg, 491002, Chhattisgarh, India}

\author{Satyajit Puhan\orcid{https://orcid.org/0009-0004-9766-5005}}
\email{puhansatyajit@gmail.com}
\affiliation{Department of Physics, Dr. B.R. Ambedkar National Institute of Technology, Jalandhar, 144008, India}
    
\author{Sabyasachi Ghosh\orcid{https://orcid.org/0000-0003-1212-824X}}
\email{sabya@iitbhilai.ac.in}
\affiliation{Department of Physics, Indian Institute of Technology Bhilai, Kutelabhata, Durg, 491002, Chhattisgarh, India}

\author{Harleen Dahiya\orcid{https://orcid.org/0000-0002-3288-2250}}
\email{dahiyah@nitj.ac.in}
\affiliation{Department of Physics, Dr. B.R. Ambedkar National Institute of Technology, Jalandhar, 144008, India}

    \begin{abstract}
We calculate the in-medium valence quark distribution of the pion immersed in a finite baryon density using the light-cone quark model. The medium-modified pion properties are obtained by using the constituent quark mass-dependent light cone wave functions. To obtain the constituent quark mass at finite baryon density, we employ the two-flavor Nambu--Jona-Lasinio model. We primarily focus on the in-medium electromagnetic form factor, distribution amplitude, and the parton distribution function of the pion. The parton distribution functions are also evolved from the model scale to a perturbative scale using next to leading order Dokshitzer–Gribov–Lipatov–Altarelli–Parisi evolution equations. Furthermore, our calculated form factors are compared with available experimental measurements and lattice quantum chromodynamics studies. We also examine the Mellin moments derived from our parton distribution functions in comparison with existing extractions and theoretical model predictions. 
    \end{abstract}
\maketitle
\section{Introduction}
The determination of hadronic properties from the distributions of their constituent partons has long been a central pursuit in nuclear and particle physics~\cite{Polchinski:2002jw,Altarelli:1977zs,Lai:2010vv,Diehl:2015uka,Belitsky:2005qn,Garcon:2002jb,Diehl:2003ny}. Experimental investigations of hadron structure have been primarily carried out through semi-exclusive and inclusive processes~\cite{JeffersonLabHallA:1999epl,Amoroso:2022eow}, while the forthcoming Electron–Ion Collider (EIC) aims to provide unprecedented insight into the partonic structure of hadrons and nuclei~\cite{AbdulKhalek:2021gbh}. On the theoretical side, numerous studies have explored hadronic properties in vacuum within a variety of effective frameworks~\cite{Bentz:1999gx,Mineo:1999eq,Mineo:2002bg,Brodsky:2014yha,deTeramond:2008ht,Puhan:2023hio,Puhan:2023ekt,Acharyya:2024enp,Sharma:2024lal,Acharyya:2024tql,Puhan:2024jaw,Puhan:2025pfs,Kaur:2020emh,Kaur:2020vkq,Kaur:2019jow,Kaur:2018dns,Lan:2019vui,Lan:2019rba,Chakrabarti:2013gra,More:2023pcy,Qian:2008px}. While such vacuum studies provide essential benchmarks, understanding the in-medium modification of hadronic properties is equally crucial. This importance arises from at least two considerations. First, medium effects on partonic distributions have been experimentally established. The nuclear structure functions measured in the EMC~\cite{EuropeanMuon:1983wih} and SLAC~\cite{Arnold:1983mw} experiments differ from those of free nucleons, highlighting the role of nuclear binding and composition. Likewise, the pion decay constant has been observed to decrease relative to its vacuum value in deeply bound pionic atom measurements~\cite{Suzuki:2002ae} and pion–nucleus scattering experiments~\cite{Friedman:2004jh}. Second, in relativistic heavy-ion collisions, hadrons formed at the later stages of the evolution propagate through a medium of finite temperature and baryon density. Their decay and transport properties thus depend on medium-modified parton distributions. In this context, in-medium modifications of bound states have also been investigated from the heavy-ion perspective~\cite{Sharma:2009hn,Zhao:2023nrz}. 
 
 The internal structure of hadrons arises from their constituent valence quarks, sea quarks, and gluons, whose distributions are encoded in the distribution amplitudes (DAs), electromagnetic form factors (EMFFs), and parton distribution functions (PDFs). The pion, being the lightest meson and a Goldstone boson of chiral symmetry breaking, has been extensively studied in this context. Numerous investigations have explored its DA~\cite{Radyushkin:2009zg,Polyakov:2009je,Zhang:2017bzy,Xu:2018eii}, EMFF~\cite{Roberts:1994hh,Ananthanarayan:2018nyx,Ydrefors:2021dwa,Chang:2013nia}, and PDF~\cite{Lan:2019rba,CHANG2015547,Zhang:2018nsy,Shi:2018mcb,Chen:2016sno} from both theoretical~\cite{Radyushkin:2009zg,Polyakov:2009je,Zhang:2017bzy,Shi:2018mcb,Xu:2018eii,Chen:2016sno,Roberts:1994hh,Ananthanarayan:2018nyx,Ydrefors:2021dwa,Chang:2013nia,Lan:2019rba,CHANG2015547,Zhang:2018nsy} and experimental~\cite{AMENDOLIA1986168,BARKOV1985365,JeffersonLabFpi:2000nlc,Wijesooriya:2005ir,BaBar:2009rrj} perspectives. The pion DA plays a central role in hard exclusive processes, as it encodes the overlap between the pion state and its valence quark–antiquark Fock component, describing how the pion’s longitudinal momentum is shared between the quark and antiquark. Its normalization yields the pion decay constant, making it a key quantity in hadronic physics. The EMFF, by contrast, characterizes the response of the pion to an electromagnetic probe and provides direct access to its internal charge distribution, including the extraction of the charge radius. The PDF describes the probability of finding a parton carrying a given longitudinal momentum fraction and can be determined from hard inclusive processes such as deep inelastic scattering. The in-medium behavior of these quantities—DA, EMFF, and PDF—has also been studied within various theoretical frameworks~\cite{Puhan:2025ibn,Kaur:2024wze,Puhan:2024xdq,deMelo:2014gea,deMelo:2016uwj,Hutauruk:2019ipp,Hutauruk:2021kej,Arifi:2023jfe,Arifi:2024tix,Gifari:2024ssz,Suzuki:1995vr}. In particular, Refs.~\cite{Puhan:2025ibn,Kaur:2024wze,Puhan:2024xdq} employed the light-cone quark model (LCQM) with constituent quark masses determined from the chiral SU(3) quark mean-field model. Studies in Refs.~\cite{deMelo:2014gea,deMelo:2016uwj} and~\cite{Arifi:2023jfe,Arifi:2024tix} utilized the Bethe–Salpeter amplitudes and Gaussian wave functions within the quark–meson coupling (QMC) model, respectively. A hybrid quark–meson coupling and Nambu--Jona-Lasinio (QMC–NJL) model approach was adopted in Refs.~\cite{Hutauruk:2018qku,Hutauruk:2019ipp}, whereas Refs.~\cite{Hutauruk:2021kej,Gifari:2024ssz} investigated the pion’s in-medium structure using the NJL–Bethe–Salpeter equation (NJL–BSE) framework.

In the present work, we investigate the valence quark distribution of the pion following an approach similar to Refs.~\cite{Puhan:2025ibn,Kaur:2024wze,Puhan:2024xdq}, which we call a hybrid light-cone quark model–Nambu–Jona-Lasinio model (LCQM–NJL) framework. In this scheme, the pion properties are described within the LCQM, while the in-medium constituent quark mass is obtained from the two-flavor NJL model in symmetric nuclear medium~\cite{Bentz:2001vc}. The LCQM effectively incorporates relativistic effects of quarks and gluons inside hadrons and is particularly suitable for studying their properties in the low-momentum, non-perturbative regime of quantum chromodynamics (QCD), where a meson is represented as a superposition of Fock states, $$\ket{\mathcal{M}}=\ket{q\bar{q}}+\ket{q\bar{q}g}+\ket{q\bar{q}q\bar{q}}+\dots,$$ with the leading component corresponding to the valence quark–antiquark configuration. The LCQM has successfully reproduced the electromagnetic form factor (EMFF) of the pion in agreement with experimental data~\cite{Ma:1992sn,Cao:1997sx} and has also been used to predict the pion charge radius and decay constant. Given its predictive capability, extending the LCQM to finite-density environments provides a natural means to explore in-medium modifications of pion properties. At finite baryon density or temperature, partial restoration of chiral symmetry occurs, and several effective QCD models—including the chiral quark model, quark–meson model, linear sigma model, and the NJL model—have been employed to study this phenomenon. 

The NJL model, in particular, has been widely used to describe the properties of quark matter and the quark–gluon plasma (QGP) created in heavy-ion collisions. Its Lagrangian captures quark interactions via four-fermion contact terms and embodies spontaneous chiral symmetry breaking, leading to dynamical mass generation through non-vanishing chiral condensates~\cite{Vogl:1991qt,Klevansky:1992qe,HATSUDA1994221,Buballa:2003qv}. The constituent quark masses are determined self-consistently from the gap equation derived from the quark self-energy. The NJL model is most effective in the low-temperature and intermediate-density regimes, where perturbative QCD becomes unreliable due to strong coupling and lattice QCD calculations are hindered by the sign problem. In the present work, we employ the two-flavor NJL model specifically formulated for nuclear matter~\cite{Bentz:2001vc}. Since our primary interest lies in understanding the effects of the nuclear medium on the constituent quark mass, the model must reproduce the properties of a free nucleon, such as its vacuum mass, in the zero-density limit while simultaneously describing the saturation properties of nuclear matter at finite baryon density. This framework was successfully developed in the seminal work of Bentz and Thomas~\cite{Bentz:2001vc}. In this approach, nucleons are described as quark--diquark bound states propagating in the background of the constituent quark vacuum. The density-dependent constituent quark mass in nuclear matter is obtained by minimizing the appropriate free energy (or grand potential), which includes contributions from vacuum quark loops, mean scalar and vector fields, and the nucleonic Fermi sea~\cite{Bentz:2002um}. The resulting constituent quark mass at zero temperature and finite baryon density is then used to construct the light-cone wave functions (LCWFs) for the valence sector of the pion.
We subsequently investigate the effect of baryon density on the pion’s distribution amplitude (DA), decay constant, EMFF, charge radius, parton distribution function (PDF), and Mellin moments, and evolve the in-medium PDFs and Mellin moments from the model scale (0.20~GeV$^{2}$) to the perturbative scale (25~GeV$^{2}$) using the next to leading order Dokshitzer–Gribov–Lipatov–Altarelli–Parisi (NLO DGLAP) evolution equations~\cite{Miyama:1995bd,Hirai:1997gb,Hirai:1997mm,Hirai:2011si}.

The present work is organized as follows. In Sec.~\eqref{NJL}, we describe the NJL model used in the present work to obtain the in-medium constituent quark mass that has been used as input to obtain the properties of the pion in nuclear matter. The LCQM and the LCWFs are briefly addressed in Sec.~\eqref{Sec:LCQM}. In Sec.~\eqref{Sec:results}, the results of the present work are discussed and finally summarized in Sec.~\eqref{Sec:conclusion}. 

\section{Formalism}
\label{sec:Formalism}

\subsection{NJL model for symmetric nuclear matter and effective quark masses}\label{NJL}
In this section, we describe the NJL model in symmetric nuclear matter as developed in Ref.~\cite{Bentz:2001vc} and subsequently employed in numerous studies of hadronic and nuclear structure~\cite{Mineo:2003vc,Cloet:2005pp,Horikawa:2005dh,Cloet:2005rt,Cloet:2006bq,Cloet:2009qs,Cloet:2012td,Wang:2021elw,Hutauruk:2021kej,Gifari:2024ssz,Hutauruk:2025bjd}, as well as in investigations of the phases of strongly interacting matter~\cite{Bentz:2002um,Lawley:2005ru,Lawley:2006ps,Noro:2023vkx,Bentz:2025rla}. Our primary objective is to determine the baryon-density dependence of the constituent quark mass in nuclear matter. The details of the model are presented in Refs.~\cite{Bentz:2001vc,Bentz:2002um}; here, we briefly summarize only the aspects relevant for the present calculations.

The NJL model is characterized by 4-fermi contact interactions between quarks. Any such 4-fermi interaction can be Fierz symmetrized and rewritten as a linear combination of the form~\cite{Bentz:2001vc,Bentz:2002um} $\sum_{i} G_{i} (\bar{\psi}\Gamma_{i}\psi)^{2}$ where $\psi\equiv\begin{pmatrix}
\psi_{u} \\
\psi_{d}
\end{pmatrix}$ denotes the $SU(2)$ flavor quark field, $\Gamma_i$ represents matrices in Dirac, flavor, and color space, and the coupling constants $G_i$ are related to those appearing in the original interaction Lagrangian. By explicitly retaining the scalar, pseudoscalar, and vector interaction channels, the Lagrangian density of the two-flavor NJL model can be written as~\cite{Vogl:1991qt,Klevansky:1992qe,HATSUDA1994221,Buballa:2003qv,Bentz:2001vc}, 
\be
\mathcal{L}=\bar{\psi}(i\slashed{\partial}-{\mathbf m})\psi +G_{\pi}\left[(\bar{\psi}\psi)^{2}+(\bar{\psi}i\gamma_{5}\vec{\sigma}\psi)^{2}\right]-G_{\omega}(\bar{\psi}\gamma^{\mu}\psi)^{2}+\cdots~,\label{AD1}
\ee
where $\mathbf{m}\equiv\begin{pmatrix}
m_{u} \\
m_{d}
\end{pmatrix}$ is the bare mass matrix for different flavors. In this paper, we assume degenerate mass for quark flavors, \textit{i.e.}, $m_{u}=m_{d}=m$. The coupling constants in the scalar, vector channels, and the Pauli matrices are respectively denoted by $G_{\pi}$, $G_{\omega}$, and $\vec{\sigma}$. For a nuclear matter with finite baryon density $\rho_{B}$, one can rewrite the Lagrangian~\eqref{AD1} as~\cite{Bentz:2001vc},
\be
\mathcal{L}=\bar{\psi}(i\slashed{\partial}-m^{\ast}-2G_{\omega}\gamma^{\mu}\omega_{\mu})\psi-\frac{(m^{\ast}-m)^{2}}{4G_{\pi}}+ G_{\omega}\omega_{\mu}\omega^{\mu}+ \mathcal{L}_{I}~,\label{AD2}
\ee
where we define the constituent quark mass $m^{\ast}\equiv m-2G_{\pi}\langle \rho_{B}\vert \bar{\psi}\psi\vert\rho_{B}\rangle$, mean vector field $\omega^{\mu}\equiv \langle \rho_{B}\vert \bar{\psi}\gamma^{\mu}\psi\vert\rho_{B}\rangle$. $\mathcal{L}_{I}$ is the normal ordered interaction Lagrangian. To construct the nucleon as a quark-diquark bound state in the NJL model, one needs to decompose the $\mathcal{L}_{I}$ in terms of $qq$ interaction channels. For the $qq$ interaction in the scalar diquark channel, we have,
\be
\mathcal{L}_{I,s}=G_{s}(\bar{\psi}(\gamma_{5}C)\sigma_{2}\beta^{A}\bar{\psi}^{T})(\psi^{T}(C^{-1}\gamma_{5})\sigma_{2}\beta^{A}\psi),\label{AD3}
\ee
where $\beta^{A}=\sqrt{\frac{3}{2}}\lambda^{A}~ (A=2,5,7)$ are the color $\bar{3}$ matrices, $C=i \gamma_{2}\gamma_{0}$ and $G_{s}$ is the coupling constant in the scalar diquark channel. The mass of the diquark is obtained from the pole of the corresponding $T$-matrix~\cite{Bentz:2001vc},
\be
\tau_{s}(q)&=&\frac{4iG_{s}}{1+2G_{s}\Pi_{s}(q)} \text{ with the } qq \text{ bubble graph (diquark polarization),}\nonumber\\
\Pi_{s}(q) &=& 6i \int \frac{d^{4}k}{(2\pi)^{4}} \tr_{\rm D} [\gamma_{5}S(k)\gamma_{5}S(-(q-k))]~,\label{AD4}
\ee
where $S(k)=\frac{1}{\slashed{k}-m^{\ast}+i\epsilon}$ is the dressed quark propagator. In the static approximation~\cite{BUCK199229,Bentz:2001vc}, the $T$-matrix for the nucleon is given by~\cite{Bentz:2001vc} 
\be
T(q)&=&\frac{3}{m^{\ast}}\frac{1}{1+\frac{3}{m^{\ast}}\Pi_{N}(q)} \text{ with the quark-diquark bubble graph (nucleon polarization) contribution,}\nonumber\\
\Pi_{N}(q) &=& - \int \frac{d^{4}k}{(2\pi)^{4}} [S(k)\tau_{s}(q-k)]~.\label{AD5}
\ee
The mass of the diquark $m_{D}=m_{D}(m^{\ast})$ and nucleon $m_{N}=m_{N}(m^{\ast})$ can be obtained from the poles of the corresponding $T$-matrices, i.e., $1+2G_{s}\Pi_{s}(q^{2}=m_{D}^{2})=0$ and $1+\frac{3}{m^{\ast}}\Pi_{N}(\slashed{q}=m_{N})=0$. It is well known that the NJL Lagrangian is non-renormalizable~\cite{Vogl:1991qt,Klevansky:1992qe,HATSUDA1994221,Buballa:2003qv}; therefore, a regularization scheme must be specified in order to fully define the model. In the present work, we employ the proper-time regularization method following Ref.~\cite{Bentz:2001vc}. Within this scheme, the loop integrals (e.g., Eqs.~\eqref{AD4} and \eqref{AD5}) are evaluated by first applying the Feynman parametrization to the product of propagators, followed by a Wick rotation. The resulting integrals are then regularized through the introduction of infrared ($\Lambda_{\rm IR}$) and ultraviolet ($\Lambda_{\rm UV}$) cutoffs. While the ultraviolet cutoff is required to render the loop integrals finite, the infrared cutoff incorporates certain aspects of confinement physics, as discussed in Ref.~\cite{Bentz:2001vc}. The polarization tensors (bubble-diagram contributions) for both the diquark and the nucleon within the proper-time regularization scheme are presented in Appendix~\eqref{ape1}.

To describe the symmetric nuclear matter within the NJL model, one can have a ``hybrid approximation'' for the nuclear matter ground state as described in Ref.~\cite{Bentz:2001vc}. In this approximation the expectation value of any local quark operator $\mathcal{O}$ consists of expectation value in the valence quark vacuum $\vert \rho_{B}=0 \rangle=\vert 0 \rangle$ and an average over the nucleonic Fermi sea $\vert N, \vec{k} \rangle$ consisting of correlated valence ($v$) quarks (quark-diquark states), i.e., 
\be
\langle\rho_{B}\vert\mathcal{O} \vert \rho_{B}\rangle&=&\langle 0 \vert\mathcal{O} \vert 0 \rangle + \gamma_{N}\int \frac{d^{3}\vec{k}}{(2\pi)^{3}}  \langle \mathcal{O}\rangle_{v} f \text{ where, } \label{AD6}\\
\langle \mathcal{O}\rangle_{v} &\equiv& \int d^{3}\vec{r}~[\langle N,\vec{k}\vert \mathcal{O}(\vec{r})\vert N, \vec{k} \rangle-\langle 0\vert \mathcal{O}(\vec{r}) \vert 0\rangle]~. \label{AD7}
\ee
In writing Eq.~\eqref{AD6}, we have divided a volume factor throughout. $\gamma_{N}=2\times 2=4$ is the degeneracy factor for the nucleons, and $f$ corresponds to the distribution of the nucleons (which is actually a step function at zero temperature).  
The readers are referred to the article~\cite{Bentz:2001vc} for the discussions related to this and many more interesting points. Here we write the grand canonical potential ($\Omega$) of this system and the subsequent gap equations, which can be solved to get the density-dependent quark mass. At $T\rightarrow 0$, the grand potential obtained from the Lagrangian is written as~\cite{Bentz:2002um},
\be
\Omega &=&\langle \rho_{B} \vert \mathcal{H} \vert \rho_{B}\rangle -\mu_{B}\rho_{B}\nonumber\\
&=& i\gamma_{q}\int \frac{d^{4}k}{(2\pi)^{4}} \ln \frac{k^{2}-m^{\ast 2}+i\epsilon}{k^{2}-m_{0}^{\ast 2}+i\epsilon} + \frac{(m^{\ast}-m)^{2}}{4G_{\pi}}-\frac{(m^{\ast}_{0}-m)^{2}}{4G_{\pi}}-G_{\omega}\omega^{\mu}\omega_{\mu} +\gamma_{N}\int \frac{d^{3}\vec{k}}{(2\pi)^{3}}  f \epsilon -\mu_{B} \rho_{B}~,\label{AD8}
\ee
where the first term is the vacuum contribution due to the quark loop ($\gamma_{q}=3\times 2 \times 2=12$), the second and fourth terms are the mean scalar and vector field terms, and the fifth term is the nucleonic Fermi sea contribution. In obtaining Eq.~\eqref{AD8}, additional terms have been appropriately subtracted to make the vacuum contribution vanish at zero density~\cite{Bentz:2002um}.
$\epsilon$ is the energy of the nucleon moving in the scalar and vector fields. 
For the nuclear matter at rest, the spatial component of the vector field $\omega^{i}$ vanishes.
The gap equations are obtained by minimizing the $\Omega$ with respect to $m^{\ast}$ and $\omega^{0}$~\cite{Bentz:2002um}, 
\be
&&\left(\frac{\partial \Omega}{\partial m^{\ast}}\right)_{\mu_{B},\omega^{0}}=0~,\label{AD9}\\
&&\left(\frac{\partial \Omega}{\partial \omega^{0}}\right)_{\mu_{B},m^{\ast}}=0~.\label{AD10}
\ee
These equations are simplified in the appendix~\eqref{ape2}. Here, we write down the final form of the gap equations that follow from Eqs.~\eqref{AD9} and \eqref{AD10},
\be
&& \omega_{0}=3\rho_{B}=3~\frac{\gamma_{N}}{6\pi^{2}}((\mu_{B}^{\ast})^{2}-m_{N})^{3/2}~,\label{AD11}\\
&& m^{\ast}=m+2G_{\pi} \bigg[\frac{\gamma_{q}m^{\ast}}{8\pi^{2}}\int_{1/\Lambda_{\rm UV}^{2}}^{1/\Lambda_{\rm IR}^{2}} \frac{e^{-\tau m^{\ast 2}}}{\tau^{2}} d\tau -m_{N}\frac{\partial m_{N}}{\partial m^{\ast}} \frac{\gamma_{N}}{4\pi^{2}} \left(\mu_{B}^{\ast}\sqrt{\mu_{B}^{\ast 2}-m_{N}^{2}}-m_{N}^{2}\sinh^{-1}{\frac{\sqrt{\mu_{B}^{\ast 2}-m_{N}^{2}}}{m_{N}}}\right) \bigg]\label{AD12},
\ee
where $\mu_{B}^{\ast}=\mu_{B}-6G_{\omega}\omega^{0}$. One can use these two gap equations~\eqref{AD11} and \eqref{AD12} to get the density-dependent quark mass, which are further discussed in Sec.~\eqref{conmass}.

\subsection{Light-cone quark  model}
\label{Sec:LCQM}
The eigenstate of a meson $|\mathcal{M}(P^+,\bfP,S_z)\rangle$ carrying a total four momenta $P\equiv(P^+,P^-,\bfP)$ and longitudinal spin projection $S_z$ can be written in the LCQM as follows~\cite{Brodsky:1997de,Brodsky:2014yha}:
\be
|\mathcal{M}(P^+,\bfP,S_z)\rangle &=& \sum_{n,\lambda_j} \int \prod_{j=1}^{n} \frac{dx_j~  d^2\bfkj}{2(2\pi)^3\sqrt{x_{j}}} \, 16 \pi^{3} \, \delta \bigg(1-\sum_{j=1}^{n} x_{j}\bigg) \, \delta^{(2)} \bigg(\sum_{j=1}^{n}\bfkj\bigg) \nonumber \\		
&\times& \psi_{n/\mathcal{M}}(x_{j},\bfkj,\lambda_{j})|n; x_{j} P^{+},x_{j}\bfP + \bfkj,\lambda_{j}\rangle \, ,
\label{MesonState}
\ee
where $x_j=\frac{\textbf{k}_j^+}{P^+}$, $\bfkj$ and $\lambda_j$ are, respectively, the longitudinal momentum fraction, transverse (internal) momentum, and helicity carried by the $j$th constituent parton. The delta functions occurring in Eq.~\eqref{MesonState} ensure the conservation of longitudinal and transverse momentum.
The normalization of the multiparticle state $|n\rangle$ reads as 
\be
\langle n; k^{\prime +}_j, \bfkj^\prime, \lambda_{j}^\prime|n ; k^+_j, \bfkj, \lambda_j \rangle = \prod_{j=1}^{n} 16 \pi^{3} \,  k^{\prime +}_j \, \delta (k^{\prime +}_j-k^+_j) \, \delta^{(2)} ( \bfkj^\prime-\bfkj) \, \delta_{\lambda_{j}^\prime \lambda_{j}} \, .
\ee
The $\psi_{n/\mathcal{M}}(x_{j},\bfkj,\lambda_{j})$ are the light-cone wave functions (LCWFs) of the given meson. In general, the eigenstate of a meson can have contributions from the Fock states containing valence quarks as well as sea quarks and gluons. In the following, we consider the two-particle  Fock state of the pion consisting only of the valence quarks--$u$ and $\bar{d}$,
\be 
|\Pi (P^+,\bfP,S_z)\rangle &=&\sum_{\lambda_{1},\lambda_{2}}\int \frac{dx_{1}dx_{2}d^{2}{\bf k}_{1\perp}d^{2}{\bf k}_{2\perp}}{2(2\pi)^{3}~2(2\pi)^{3}\sqrt{x_{1}x_{2}}}~2(2\pi)^{3}\delta(1-x_{1}-x_{2})~\delta^{2}(\vec{k}_{1\perp}+\vec{k}_{2\perp})\nonumber\\
&&\psi(x_{1},x_{2},{\bf k}_{1\perp},{\bf k}_{2\perp},\lambda_{1},\lambda_{2})|x_{1}P^{+},x_{2}P^{+},x_{1}\vec{P}_{\perp}+{\bf k}_{1\perp},x_{2}\vec{P}_{\perp}+{\bf k}_{2\perp},\lambda_{1},\lambda_{2}\rangle\nonumber,
\ee
where we use the kinematic variables with subscripts $1$ and $2$ for the $u$-quark and $\bar{d}$-quark, respectively. Simplifying the above integrals by integrating over $x_{2}$ and $\vec{k}_{2\perp}$ we have,  
\be
|\Pi (P^+,\bfP,S_z)\rangle &=&\int \frac{dx \, d^2 \bfk}{  16 \pi^3 \sqrt{x(1-x)}} \, \big[\psi (x,\bfk,\uparrow,\uparrow) \, |x P^+, \bfk, \uparrow, \uparrow \rangle   \nonumber \\
&+& \psi (x,\bfk,\uparrow,\downarrow) \, |x P^+, \bfk, \uparrow, \downarrow \rangle + \psi (x,\bfk,\downarrow,\uparrow) \, |x P^+, \bfk,  \downarrow,\uparrow \rangle \nonumber \\ &+& \psi (x,\bfk,\downarrow,\downarrow) \, |x P^+, \bfk, \downarrow, \downarrow \rangle \big] \, \label{eqnq} ,
\ee 
where we ignore the subscripts in the momentum labels of the $u$-quark. We also suppress the anti-quark labels inside the two-particle Fock basis $|\rangle$. Consistent with the constraint 
$x_1 + x_2 =1$, if the $u$-quark carries the $x$ fraction of the longitudinal momentum, the $\bar{d}$-quark will carry $(1-x)$ fraction of the total longitudinal momentum. Similarly, the internal part of the transverse momentum of the $u$ and $\bar{d}$ are equal in magnitude but opposite in direction. One writes the momenta of the meson and its constituent quarks as,
\be 
P &=& \bigg(P^+,\frac{M^{\ast2}}{P^+},\bfz\bigg) \, ,  \nonumber \\
k_1 &=& \bigg(x P^+,\frac{\bfk^2 + m^{\ast2}_u}{x P^+},\bfk \bigg) \, ,  \nonumber \\
k_2 &=& \bigg((1-x) P^+,\frac{\bfk^2 + m^{\ast2}_{\bar{d}}}{(1-x) P^+},-\bfk \bigg),  
\ee
where $M^{\ast2} = \frac{m_u^{\ast2} + \bfk^2}{x} + \frac{m_{\bar{d}}^{\ast2} + \bfk^2}{1-x}$. Throughout the paper, we take the in-medium quark and anti-quark mass to be the same, i.e., $m_{u}^{\ast}=m_{\bar{d}}^{\ast}=m^{\ast}$. The LCWFs for pion can be further expressed in terms of momentum space $\varphi$ and spin $\mathcal{S}$ wave functions as \cite{Choi:1996mq,Huang:1994dy} 
\be 
\psi(x,\bfk,\lambda_1, \lambda_2)= \varphi(x,\bfk) \, \mathcal{S} (x,\bfk,\lambda_1, \lambda_2) \, ,
\ee 
where $\lambda_1(\lambda_2)$ represent the helicity of the quark (anti-quark). 
The spin part of the wave function in the light front can be obtained from the instant-form wave function with the aid of Melosh-Wigner rotation \cite{Melosh:1974cu,Xiao:2002iv} or equivalently by choosing the proper quark-meson vertex~\cite{Xiao:2003wf} $\mathcal{S}=\frac{1}{\sqrt{2M^{\ast}}}\bar{u}(xP^{+},\bfk,\lambda_{1})~\gamma_{5}~v((1-x)P^{+},-\bfk,\lambda_{2})$,
\be 
\mathcal{S}(x,\bfk,\uparrow,\uparrow)=-\frac{k_{x}-ik_{y}}{\sqrt{2(m^{\ast 2}+\bfk^{2})}} , \nonumber \\
\mathcal{S}(x,\bfk,\uparrow,\downarrow)=\frac{m^{\ast}}{\sqrt{2(m^{\ast 2}+\bfk^{2})}} , \nonumber \\
\mathcal{S}(x,\bfk,\downarrow,\uparrow)=-\frac{m^{\ast}}{\sqrt{2(m^{\ast 2}+\bfk^{2})}} , \nonumber \\
\mathcal{S}(x,\bfk,\downarrow,\downarrow)=-\frac{k_{x}+ik_{y}}{\sqrt{2(m^{\ast 2}+\bfk^{2})}}. 
\label{SpinWfns}
\ee  
Adopting the Brodsky-Huang-Lepage prescription~\cite{Yu:2007hp,Xiao:2002iv,Kaur:2020vkq}, the momentum space wave function is given by,
\be 
\varphi (x,\bfk)=A \, exp \,\Biggl[-\frac{M^{\ast2}-2(m^{\ast2}_{u}+m^{\ast2}_{\bar{d}})}{8 \beta^2}\Biggr]=A \, exp \,\Biggl[\frac{m^{\ast2}}{2\beta^{2}}\Biggr]\, exp \,\Biggl[-\frac{m^{\ast2}+\bfk^{2}}{8x(1-x)\beta^{2}}\Biggr] , \label{momspace}
\ee 
where $A$ and $\beta$ are the normalization constant and harmonic scale parameter, respectively. One can observe that the spin wave functions in Eq. (\ref{SpinWfns}) satisfy the following normalization relation,
\be 
\sum_{\lambda_1 \lambda_2} \mathcal{S}^\ast (x,\bfk,\lambda_1,\lambda_2) \,\mathcal{S}(x,\bfk,\lambda_1,\lambda_2) = 1 \, .
\ee
Since the total wavefunction (space and spin) has to be normalized, this gives the following normalization condition for the space part of the wave function,
\be
\sum_{\lambda_1 \lambda_2}\int \frac{{d x} d^2 \bfk}{2 (2 \pi)^3} \psi^{\ast}(x,\bfk,\lambda_1, \lambda_2) \psi(x,\bfk,\lambda_1, \lambda_2)= \int \frac{{d x} d^2 \bfk}{2 (2 \pi)^3} \, |\varphi (x,\bfk)|^2 =1~.
\ee

\section{Results and discussion}
\label{Sec:results}
\subsection{In-Medium Constituent Quark Mass}\label{conmass}
\begin{figure}[H]
\centering
\includegraphics[width=10cm]{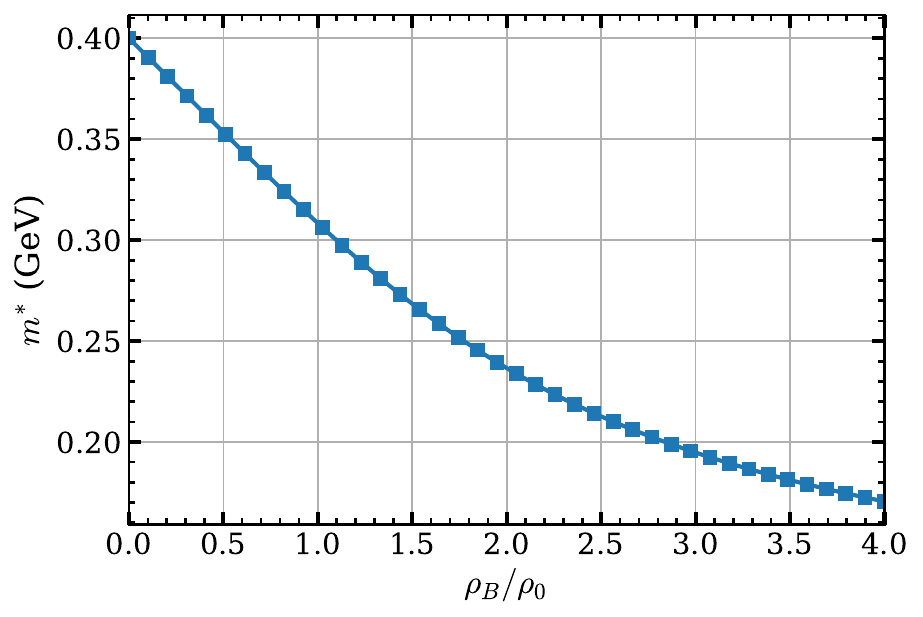}
\caption{\label{figcons} The constituent quark mass has been plotted as a function of baryonic density $\rho_B/\rho_0$. }
\end{figure}

After discussing the NJL model framework and the LCQM description of the pion state in the previous two sections, we now turn to the medium-modified pion distribution amplitude (DA), parton distribution function (PDF), and electromagnetic form factor (EMFF). A closer inspection of the LCWFs shows that, up to the valence Fock sector, the properties of the pion are primarily determined by two quantities: the constituent quark mass ($m^{\ast}$) and the harmonic scale parameter ($\beta$). In the LCQM framework, these quantities are commonly determined through variational methods by fitting meson masses to the expectation values of QCD-motivated effective Hamiltonians in vacuum~\cite{Dhiman:2019ddr}. Alternatively, they may also be constrained by comparing experimentally measured mesonic observables, such as decay constants, with their theoretical counterparts calculated using the LCWFs. For a fixed constituent quark mass, the harmonic scale parameter can be determined by requiring the pion decay constant to reproduce its experimental value, $f_{\pi}\approx130$ MeV, in vacuum. The expression for the pion decay constant in terms of the LCWFs is given in Eq.~\eqref{fpiAD}. Using the vacuum constituent quark mass, $m^{\ast}_{0}\equiv m^{\ast} (\rho_{B}=0)=0.4$ GeV, we obtain the harmonic scale parameter $\beta=0.2775$ GeV.
 In the present work, we assume that the harmonic parameter remains unchanged in the nuclear medium. Consequently, the only remaining medium-dependent input to the LCWFs is the density dependence of the constituent quark mass. This quantity is obtained by numerically solving Eq.~\eqref{AD12}. The model parameters adopted from Ref.~\cite{Bentz:2001vc} are: $G_{\pi}=19.60$ GeV$^{-2}$, $G_{\omega}=7.24$  GeV$^{-2}$, $G_{s}=9.96$ GeV$^{-2}$, $\Lambda_{\rm UV}=0.6385$ GeV, $\Lambda_{\rm IR}=0.200$ GeV and $m=0.01693$ GeV. The parameters $G_{\pi}$, $\Lambda_{\rm UV}$, $\Lambda_{\rm IR}$ and $m$ are fixed by requiring the model to reproduce the vacuum pion mass, pion decay constant, and constituent quark mass: $m_{\pi}=140$ MeV, $f_{\pi}^{\prime}=93$ MeV and $m^{\ast} (\rho_{B}=0)=0.4$ GeV. Within the NJL model, the vacuum pion mass and decay constant are obtained from the pole of the pion polarization function and the vacuum-to-pion matrix element, respectively~\cite{Gifari:2024ssz}. Note that, in the NJL framework, the pion decay constant is conventionally defined with an additional factor of $\sqrt{2}$ relative to the LCQM convention. The coupling constant $G_{s}$, which governs the interaction in the diquark channel, is fixed by requiring the free nucleon mass to satisfy $m_{N}(\rho_{B}=0)=939$ MeV. To determine the nucleon mass from the pole equation, $1+\frac{3}{m^{\ast}}\Pi_{N}(\slashed{q}_{N}=m_{N})=0$ we employ the same extrapolation procedure as used in Ref.~\cite{Bentz:2001vc}, $\frac{1}{m^{\ast}}\rightarrow \frac{1}{m_{0}^{\ast}}\frac{m_{0}^{\ast}+c}{m^{\ast}+c}$ with $c=0.7$ GeV, so that the static approximation reproduces results close to those obtained in the exact Faddeev approach. Finally, the vector coupling $G_{\omega}$ is determined by requiring the binding-energy-per-nucleon curve to pass through the empirical saturation point (0.16 fm$^{-3}$, $-15$ MeV). Although we have explicitly verified that our numerical results reproduce the corresponding results shown in Figs.~(5), (6), and (7) of Ref.~\cite{Bentz:2001vc}, we present here only those results relevant for the subsequent analysis.
In Fig.~\eqref{figcons}, we display the variation of the constituent quark mass with the baryon density of the medium. As expected, the constituent quark mass decreases with increasing baryon density and eventually becomes negligible at asymptotically large densities. This behavior is consistent with the expected restoration of chiral symmetry in dense baryonic matter. We now proceed to discuss the medium modifications of the pion DA, PDF, and EMFF.
\subsection{Distribution Amplitude (DA)}
\begin{figure*}[t]
\centering
\begin{overpic}[width=0.48\textwidth]{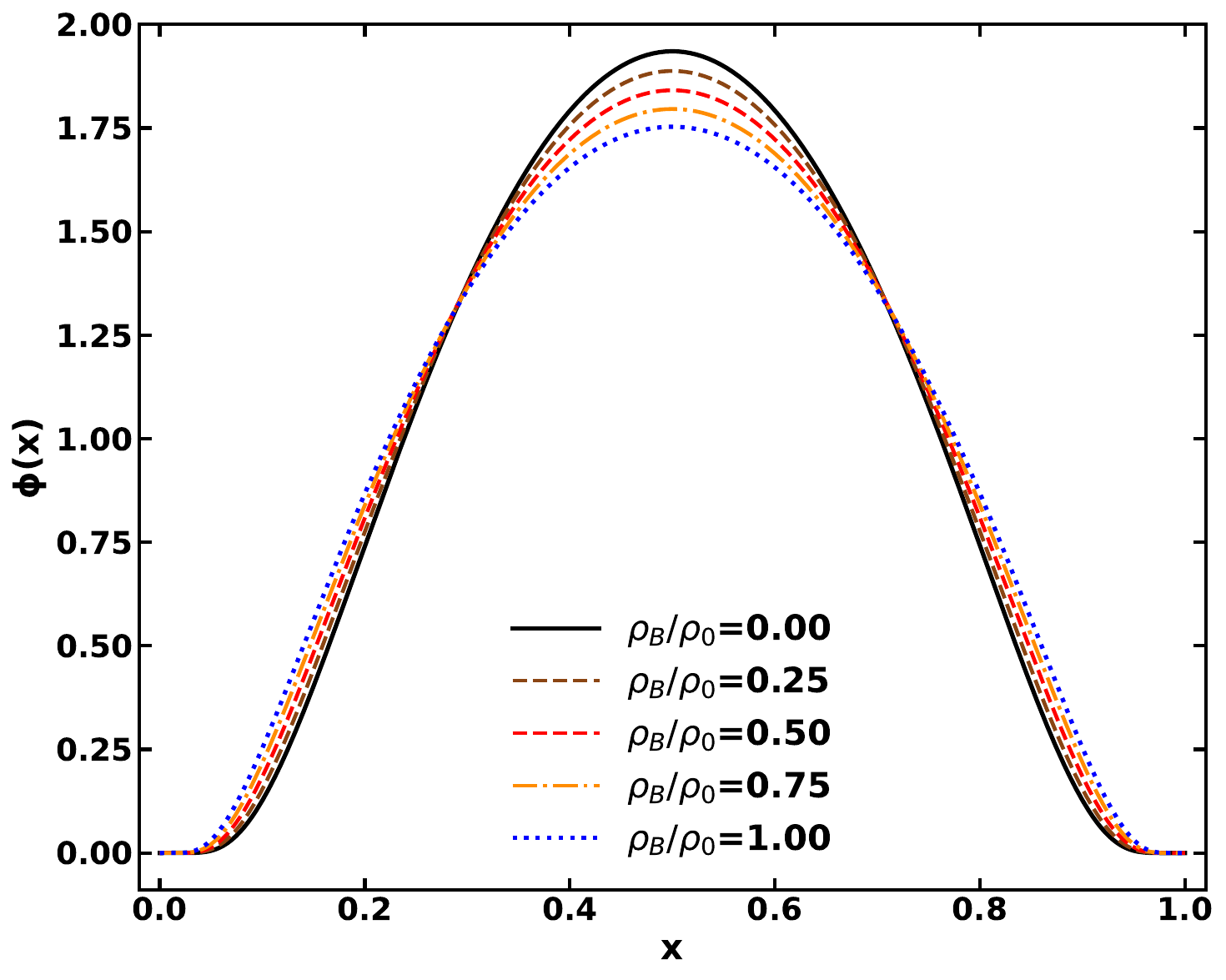}
    \put(12,62){\small (a)}
\end{overpic}
\hfill
\begin{overpic}[width=0.48\textwidth]{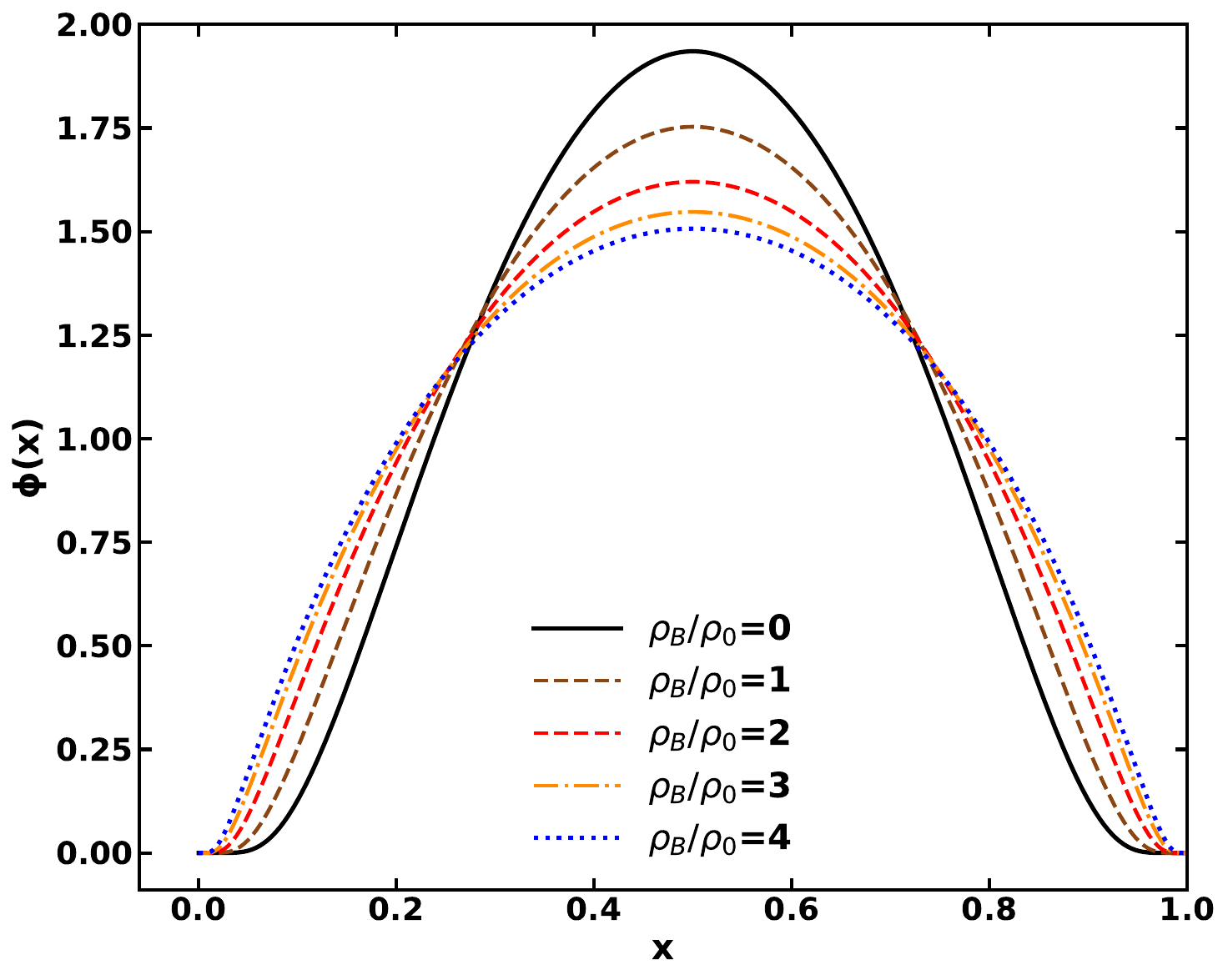}
    \put(12,62){\small (b)}
\end{overpic}

\begin{overpic}[width=0.48\textwidth]{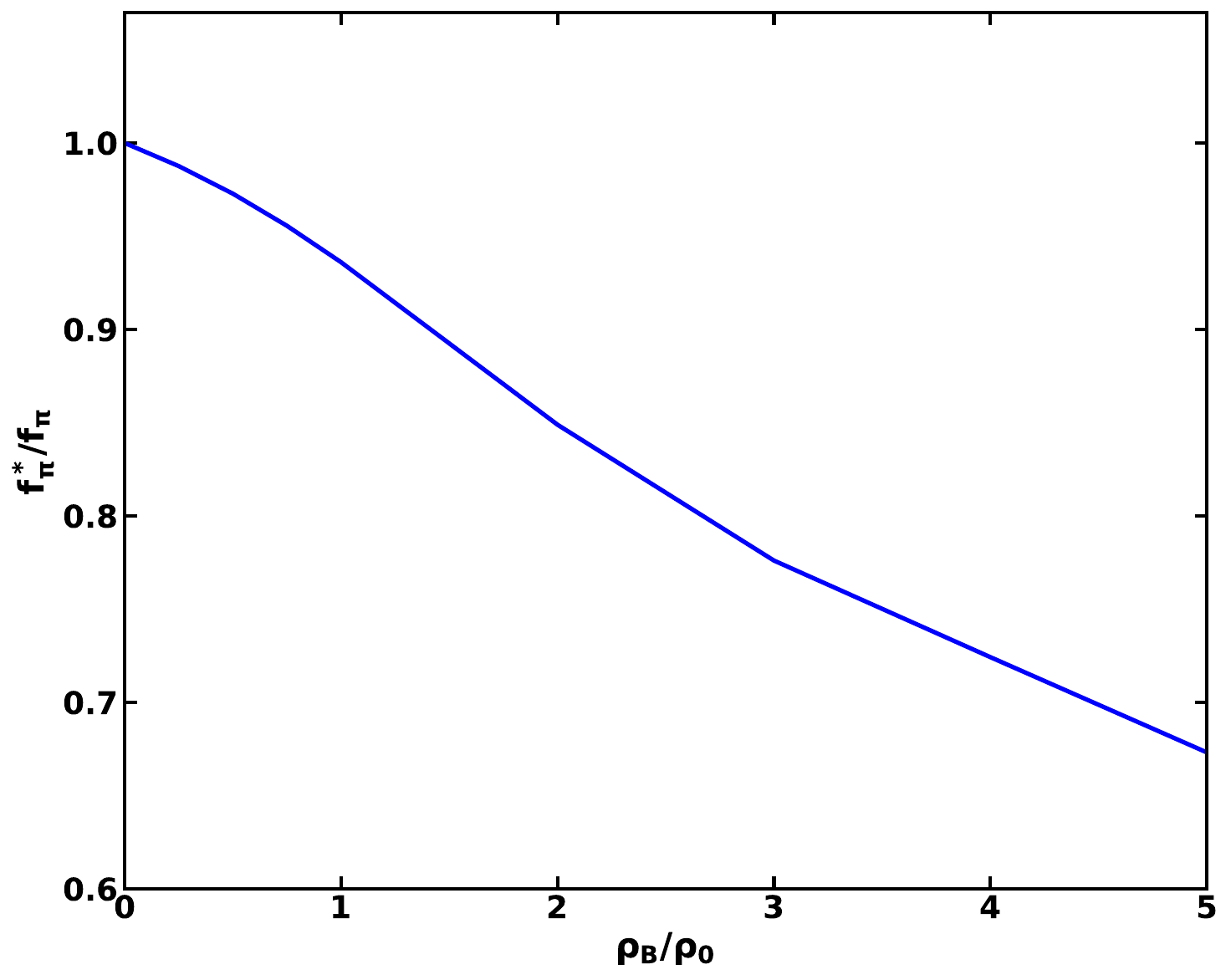}
    \put(12,62){\small (c)}
\end{overpic}
\caption{(Color online) The DA has been plotted with respect to baryonic density up to $\rho_B/\rho_0=1$ in the interval of $0.25$ in Fig. (a) and up to baryonic density  $\rho_B/\rho_0=4$ in the interval of $1$ in Fig. (b). In Fig. (c), we have plotted the ratio of the in-medium decay constant and the vacuum decay constant with respect to the baryonic density $\rho_B/\rho_0$. Here all the results are at model scale, i.e, 0.20 GeV$^2$.}
\label{fig1}
\end{figure*}

DA of the pion describes longitudinal momentum sharing between the quark and the anti-quark in its valence Fock state. The information about DAs can be extracted from the hard exclusive processes involving the pion.
In terms of the quark field correlators, it is expressed as the vacuum-to-pion matrix element as
\be
\bra{0}\bar{\vartheta}_{d}(z)i\gamma^{+}\gamma^{5}\vartheta_{u}(-z)\ket{\Pi(P^{+},{\bf P}_{\perp})}=iP^{+}f^{\ast}_{\pi}\int dx~e^{i(x-1/2)P^{+}z^{-}}\phi_{\rm DA}(x)\vert_{z^{+}={\bf z}_{\perp}=0}~,\label{DA1}
\ee
where $\vartheta$, $z=(z^{+},z^{-},{\bf z}_{\perp})$, and $f_{\pi}^{\ast}$ are the quark field operator, position four vector, and in-medium decay constant. Substitution of the free field operators leads us to the following expression for the correlators,
\be
\bra{0}\bar{\vartheta}_{d}(z)i\gamma^{+}\gamma^{5}\vartheta_{u}(-z)\ket{\Pi(P^{+},{\bf P}_{\perp})}=2\sqrt{2}iP^{+}\int \frac{dx~d^{2}{\bf k}_{\perp}}{2(2\pi)^{3}\sqrt{2x(1-x)}}~e^{i(x-1/2)P^{+}z^{-}} (\psi_{\downarrow\uparrow}-\psi_{\uparrow\downarrow})~.\label{DA2}
\ee
Introducing the color factor $\sqrt{N_{c}}$ in the above equation and equating it with  Eq.~\eqref{DA1} we have,
\be
\frac{f_{\pi}^{\ast}}{2\sqrt{2N_{c}}}\phi_{\rm DA}(x)=\int \frac{d^{2}\bfk}{2(2\pi)^{3}\sqrt{2x(1-x)}}(\psi_{\uparrow\downarrow}-\psi_{\downarrow\uparrow}).\label{fpiAD}
\ee
The DA is normalized as follows, $\int \phi_{\rm DA}(x)~dx=1$, from which one can get the in-medium pion decay constant $f_{\pi}^{\ast}$. In Fig.~\eqref{fig1}(a) and (b), we display the variation of the DA with respect to the longitudinal momentum fraction $x$. In comparison to the vacuum case, we observe that the DA gets flattened (spread out) at non-zero baryon density. Fig.~\eqref{fig1}(a) represents the variation of in-medium DA at lower baryon density $\rho_{B}=0-1 \rho_{0}$ where as Fig.~\eqref{fig1}(b) displays the same but at higher baryon density $\rho_{B}=0-4 \rho_{0}$. At increasing baryon density, we see the universal feature of suppression at mid x and increment at low and high x as observed in~\cite{Puhan:2024xdq}. Next in Fig.~\eqref{fig1}(c), we show the ratio of the in-medium decay constant to the vacuum decay constant of the pion. The results show a monotonic decrease in the decay constant of the pion with increasing baryon density. Even at the nuclear saturation density, we find the in-medium decay constant $f_{\pi}^{\ast}$ to be approximately 95\% of the vacuum value $f_{\pi}$. Similar observations have also been made in Refs.~\cite{Puhan:2024xdq,deMelo:2016uwj,Gifari:2024ssz}. The expected value of longitudinal momentum, so-called $\xi$-moment is defined as
\begin{eqnarray}
    \langle \xi^n \rangle=\int_{0}^{1} dx \xi^n \phi_{\rm DA}(x),
    \label{moment}
\end{eqnarray}
with $\xi=(1-x)-x=1-2 x$. The inverse moment $\langle x^{-1} \rangle$ can also be calculated as 
\begin{eqnarray}
    \langle x^{-1} \rangle =\int_{0}^{1} dx \frac{ \phi_{\rm DA}(x)}{x}.
\end{eqnarray}
\begin{figure}[H]
\centering
\begin{overpic}[width=0.48\textwidth]{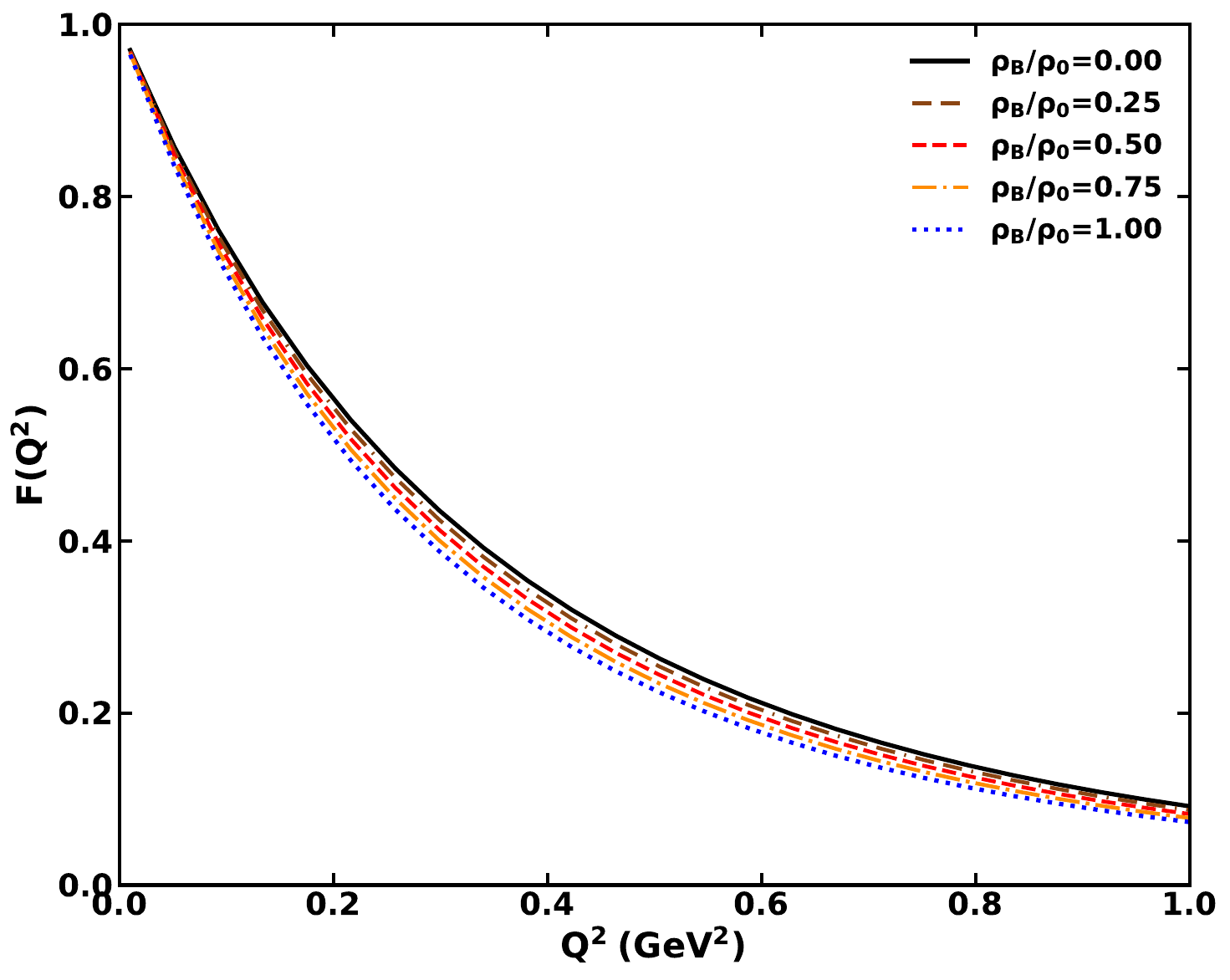}
    \put(12,58){\small (a)}
\end{overpic}
\hfill
\begin{overpic}[width=0.48\textwidth]{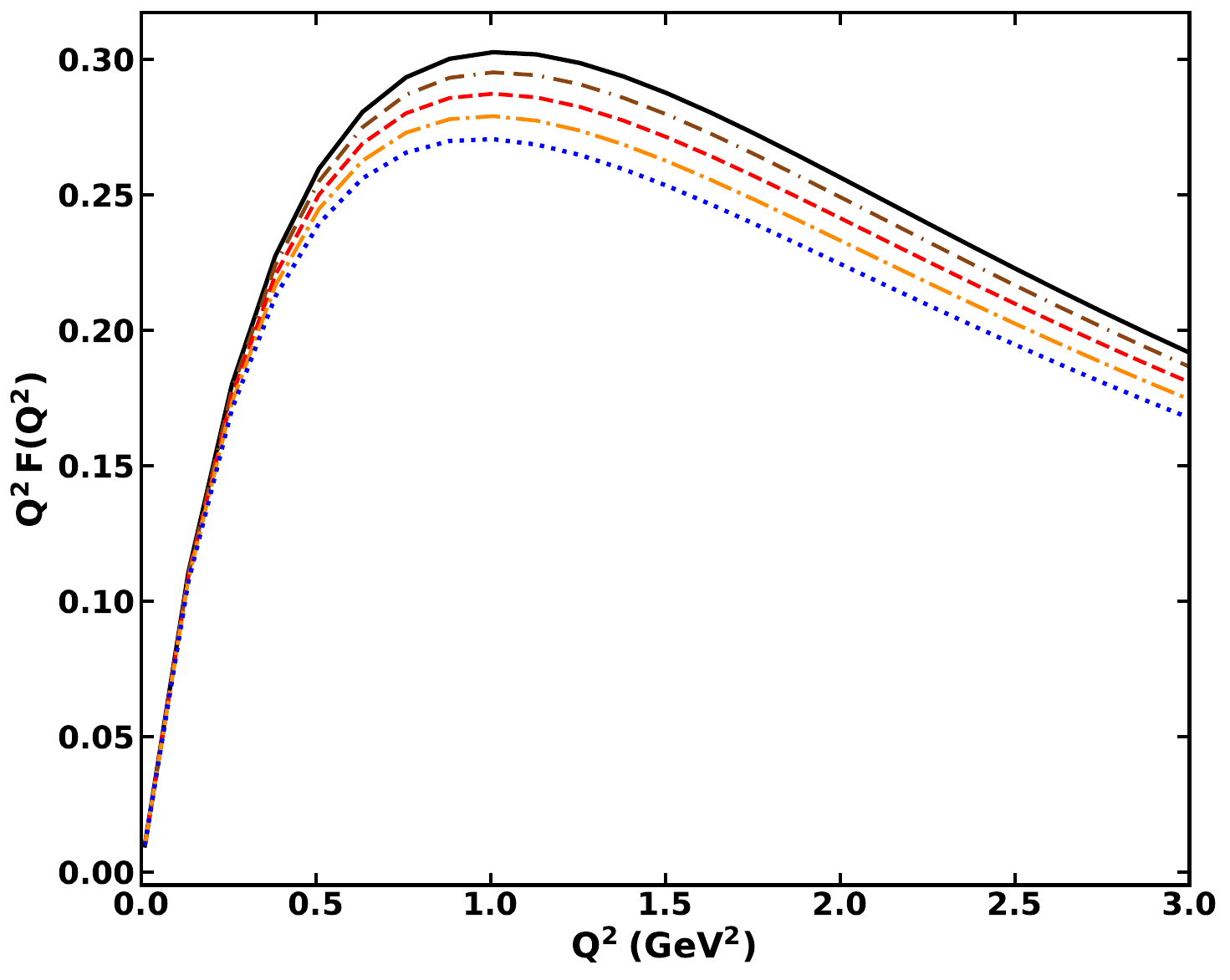}
    \put(16,58){\small (b)}
\end{overpic}

\begin{overpic}[width=0.48\textwidth]{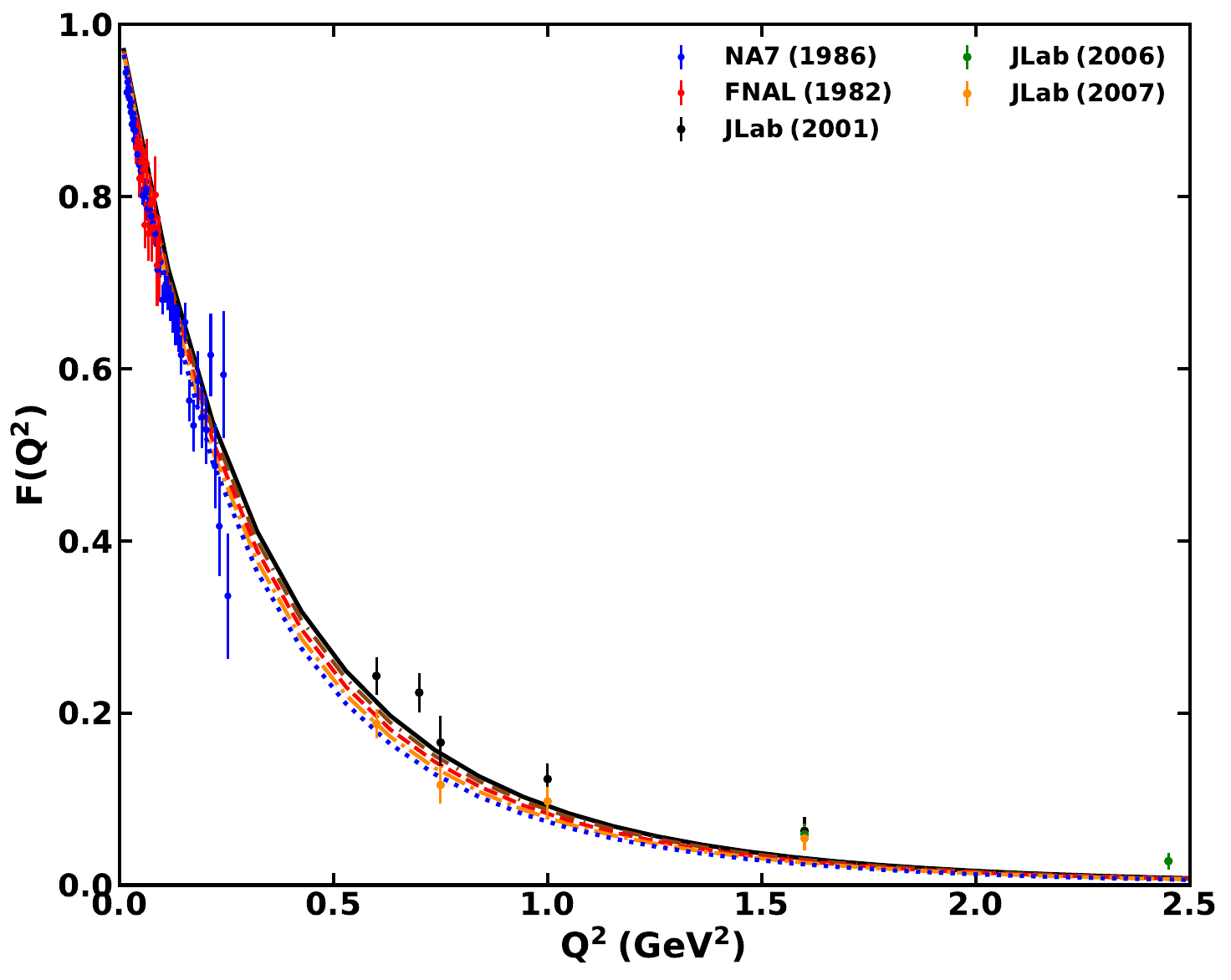}
    \put(16,58){\small (c)}
\end{overpic}
\hfill
\begin{overpic}[width=0.48\textwidth]{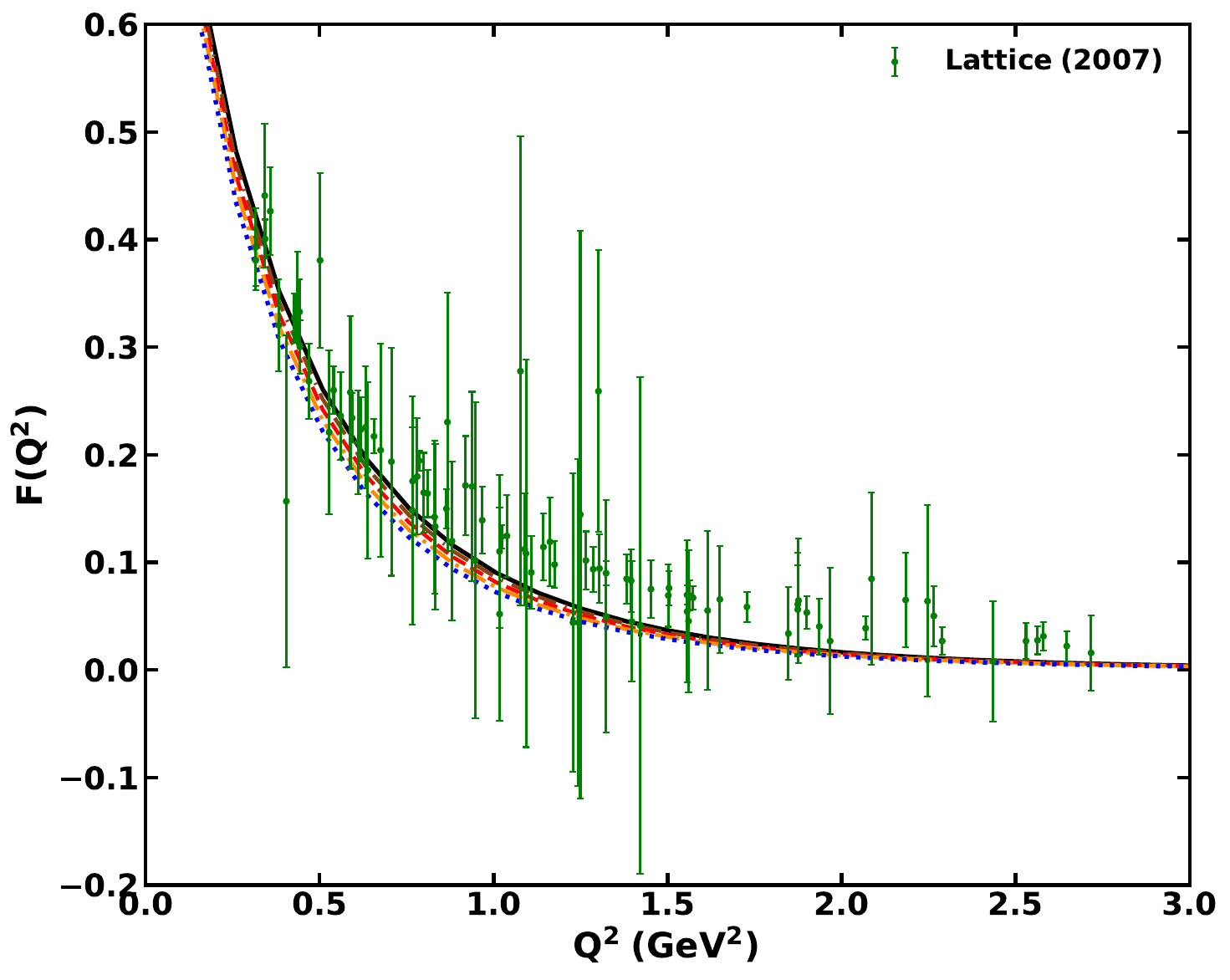}
    \put(14,58){\small (d)}
\end{overpic}

\begin{overpic}[width=0.48\textwidth]{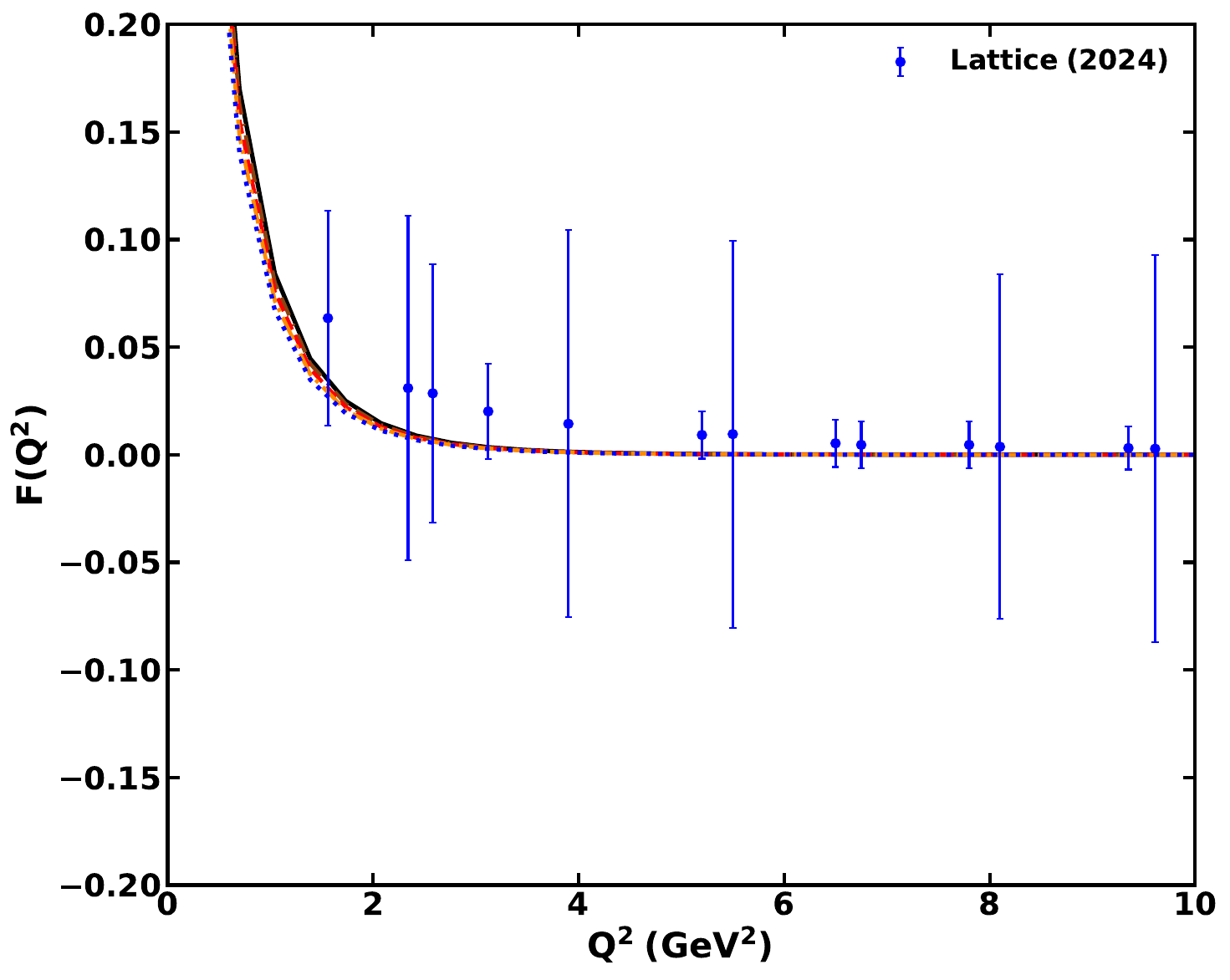}
    \put(16,58){\small (e)}
\end{overpic}
\caption{(Color online) The electro-magnetic form factor of pion has been plotted with respect to $Q^2$ GeV$^2$ at different baryonic densities up to $\rho_B/\rho_0=1$ in Fig. (a) and (b). Both the in-medium and vacuum FFs have been compared with available vacuum experimental data \cite{NA7:1986vav,JeffersonLabFpi:2007vir,JeffersonLabFpi-2:2006ysh,JeffersonLabFpi:2000nlc,Dally:1982zk} in Fig. (c) along with lattice simulations \cite{QCDSFUKQCD:2006gmg,Ding:2024lfj} in Fig. (d) and (e).}
\label{fig2}
\end{figure}
These moments up to $n=8$ along with the inverse moment have been presented in Table \ref{tab1}. The $\langle \xi^n \rangle$ moment is found to be increasing with an increase in Baryonic density, and a similar kind of observation is also seen for the case of inverse moment. At free space, the inverse moment is found to be $2.385$, which is comparable with the LFHM of $2.62$ \cite{Puhan:2023ekt}, the Light-front constituent model of $2.82$ \cite{Lorce:2016ugb}, and the LFQM of 3.11 \cite{Choi:2024ptc}. 
\begin{table}
    \centering
    \begin{tabular}{|c|c|c|c|c|c|c|}
    \hline
        $\rho_B/\rho_0$ & $\langle \xi^2 \rangle$& $\langle \xi^4 \rangle$& $\langle \xi^6 \rangle$&  $\langle \xi^8 \rangle$ & $\langle x^{-1} \rangle$\\ \hline
        $0$ & 0.1275& 0.0367& 0.0142&0.0065 & 2.385  \\ \hline
        $1$ & 0.1498& 0.0492& 0.0214 & 0.0108 & 2.496\\ \hline
        $2$ & 0.1701& 0.0621& 0.0295&0.0162 & 2.619  \\ \hline
        $3$ & 0.1831& 0.0710& 0.0355&0.0204 & 2.713  \\ \hline
        $4$ & 0.1910& 0.0766& 0.0396&0.0233 & 2.778  \\ \hline
    \end{tabular}
\caption{The $\langle \xi^n \rangle$ moment up to $n=8$ and inverse moment have been predicted at different baryonic densities up to $\rho_B/\rho_0=4$ at the model scale. }
    \label{tab1}
\end{table}

\subsection{Electromagnetic Form Factors (EMFF)}
\begin{figure}[H]
\centering
\begin{overpic}[width=0.48\textwidth]{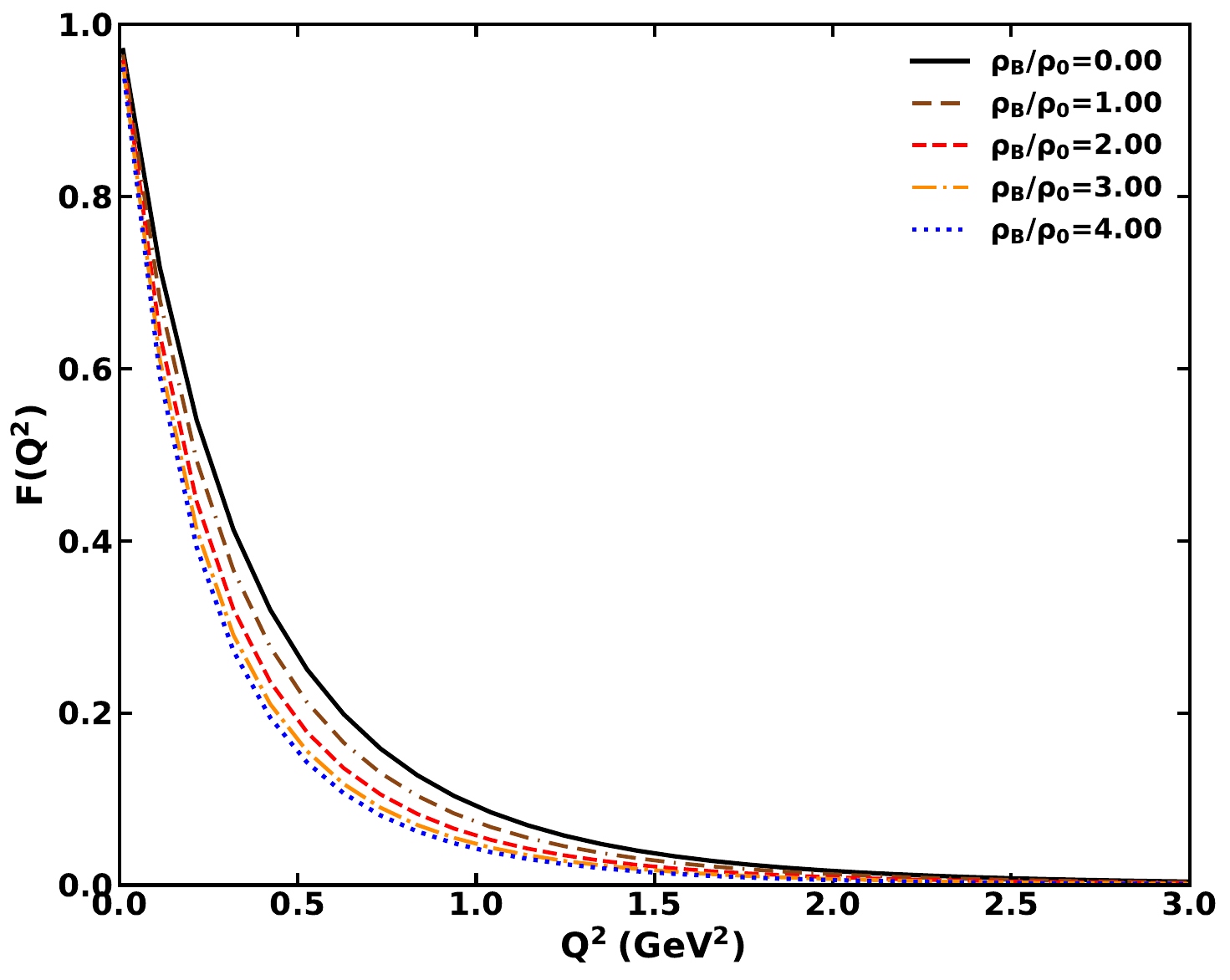}
    \put(16,58){\small (a)}
\end{overpic}
\hfill
\begin{overpic}[width=0.48\textwidth]{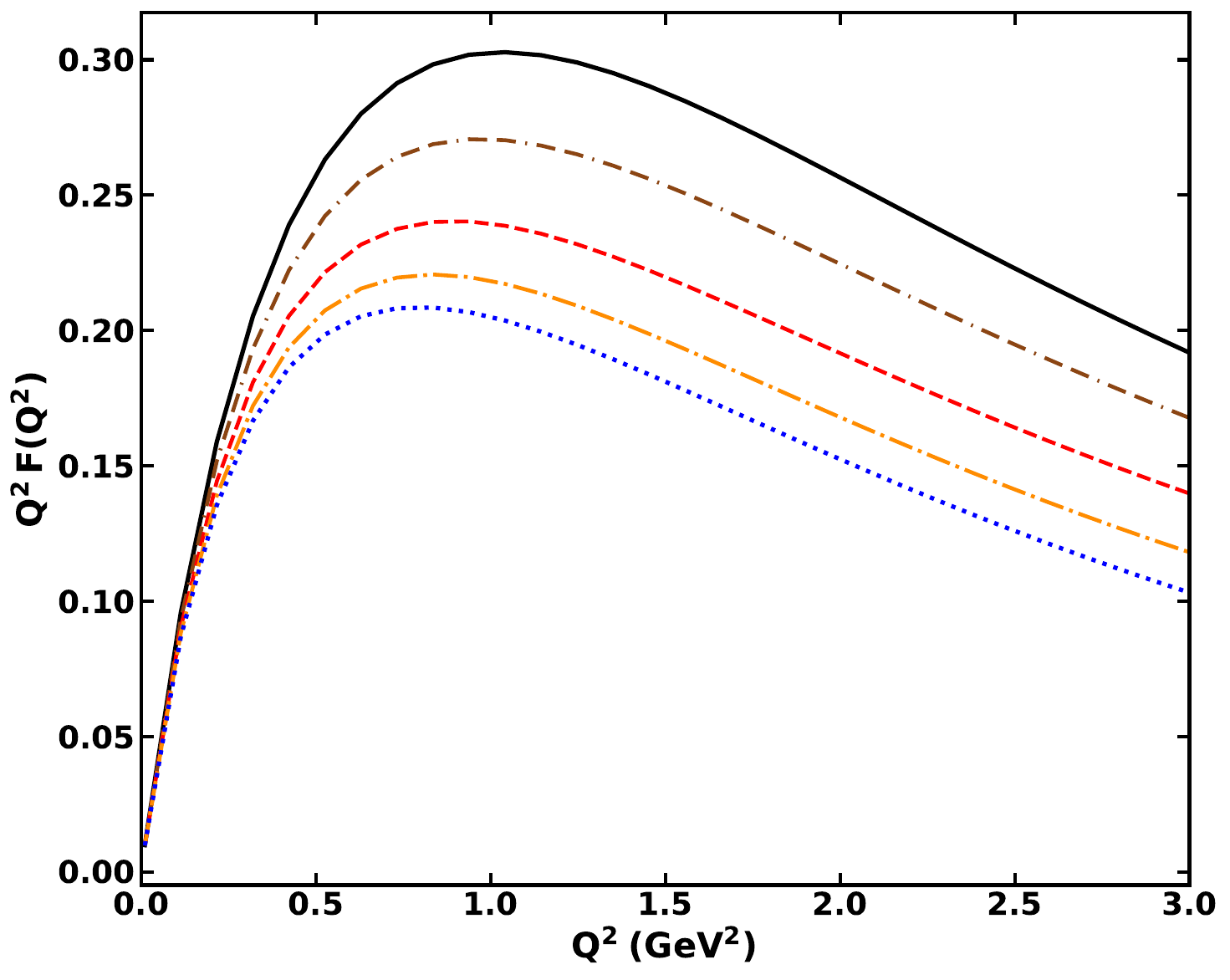}
    \put(16,58){\small (b)}
\end{overpic}

\begin{overpic}[width=0.48\textwidth]{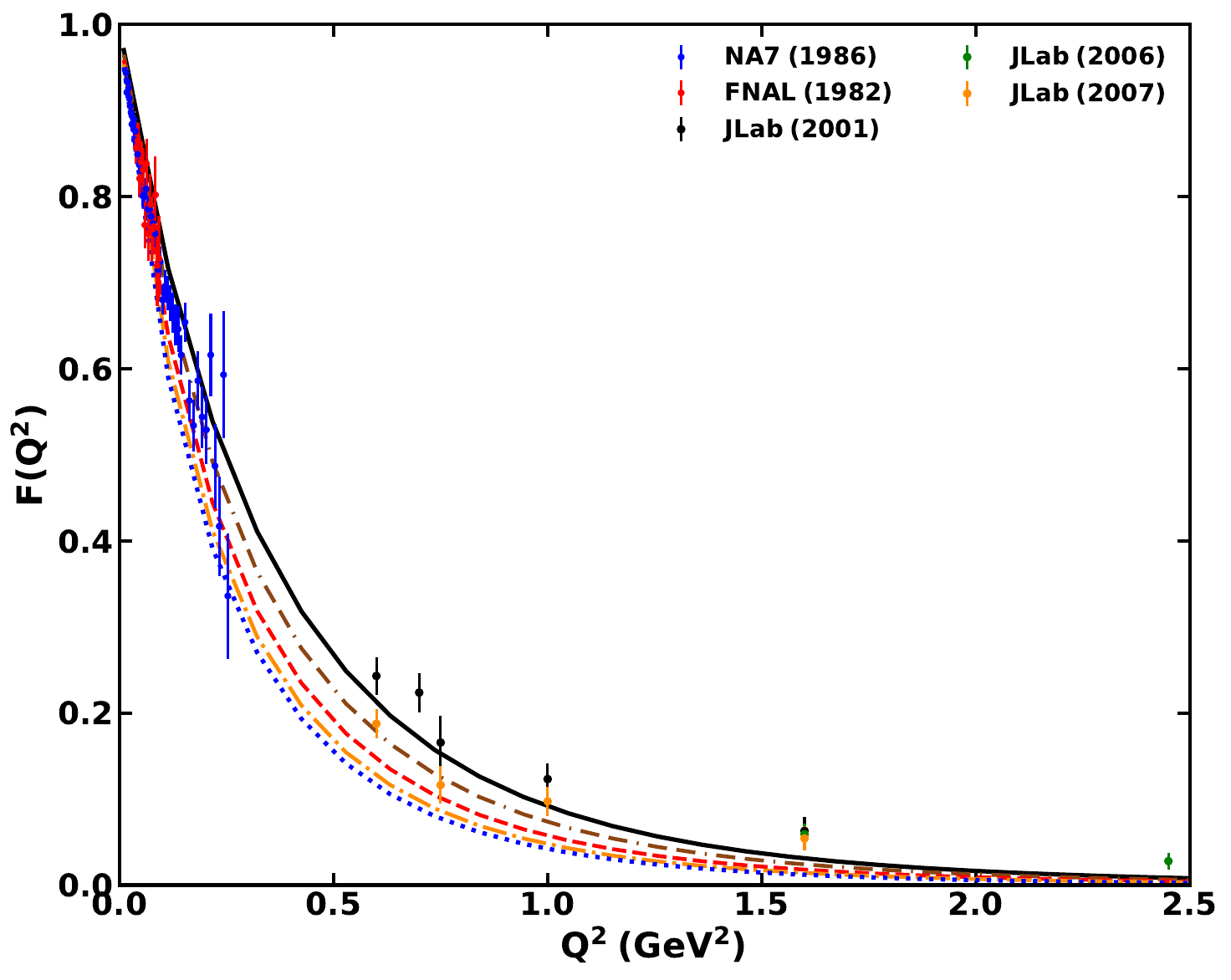}
    \put(16,58){\small (c)}
\end{overpic}
\hfill
\begin{overpic}[width=0.48\textwidth]{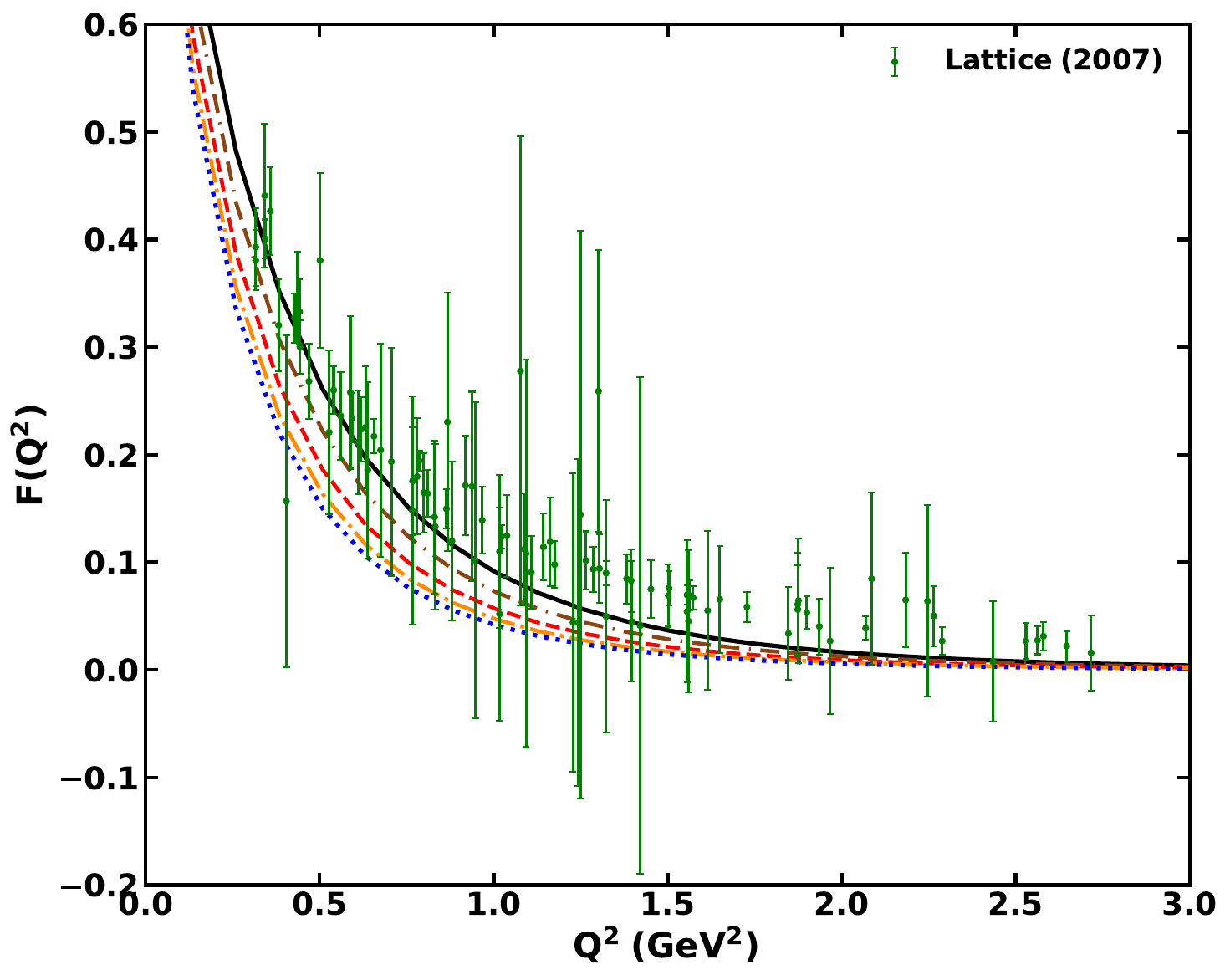}
    \put(13,58){\small (d)}
\end{overpic}

\begin{overpic}[width=0.48\textwidth]{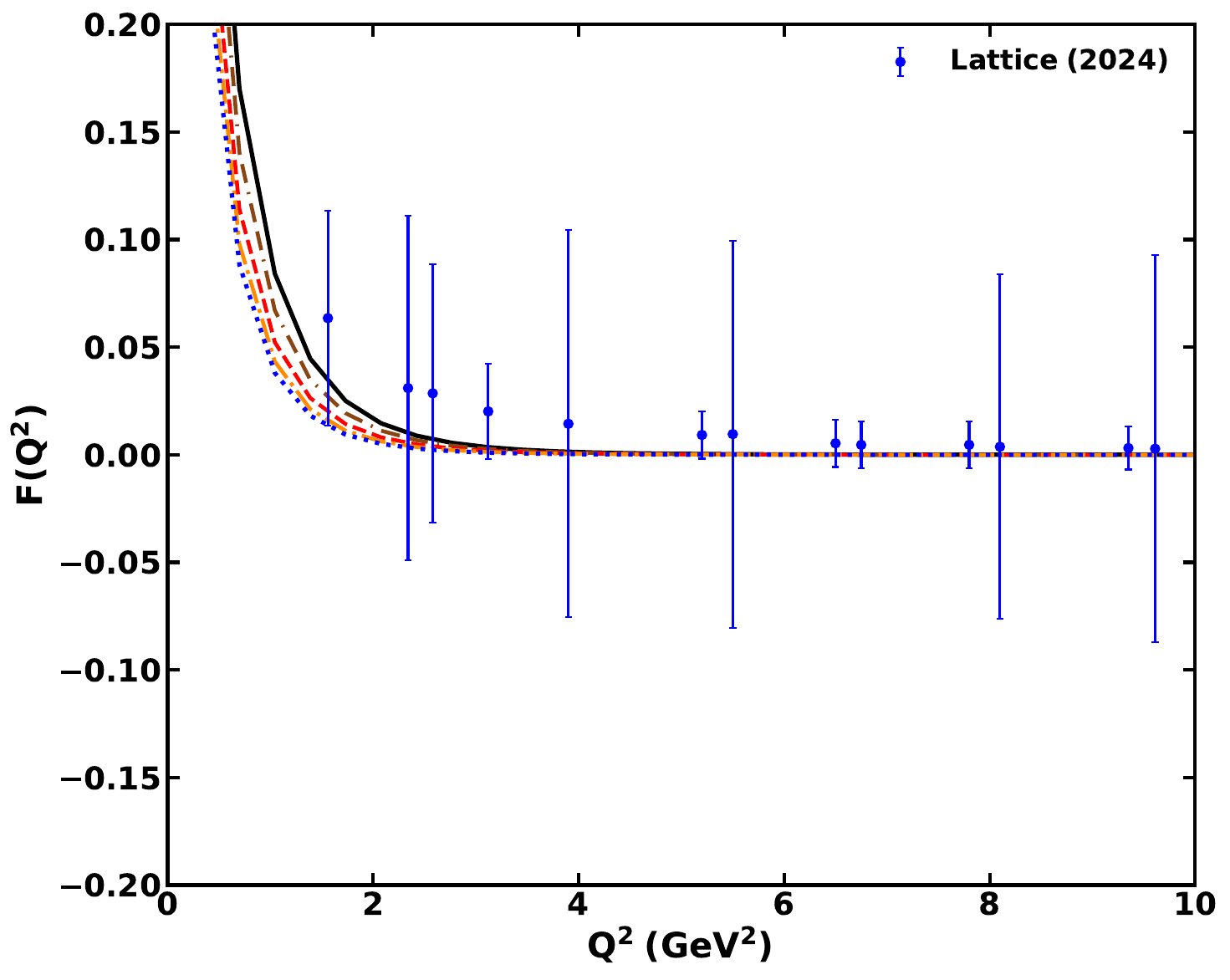}
    \put(15,58){\small (e)}
\end{overpic}
\hfill
\begin{overpic}[width=0.48\textwidth]{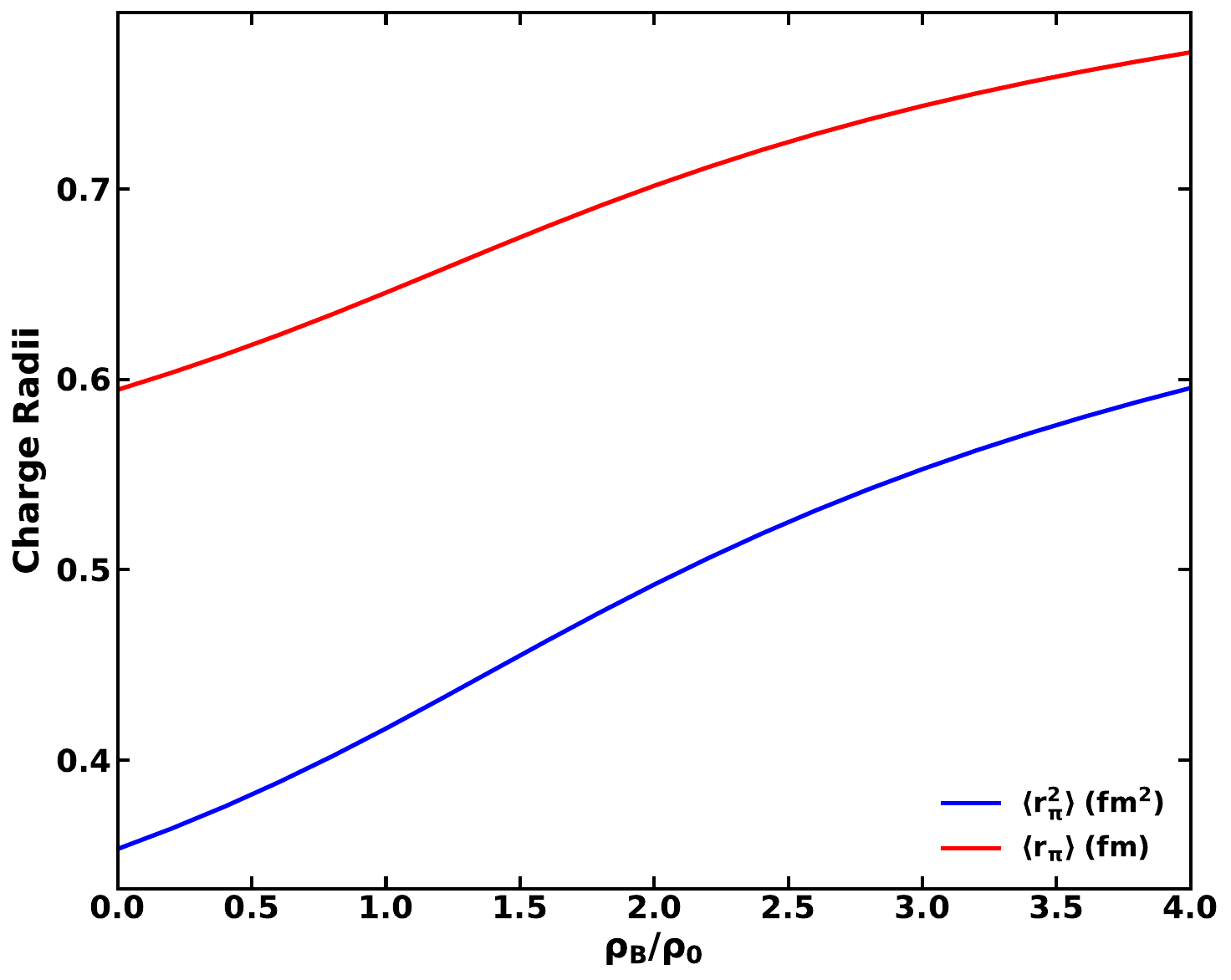}
    \put(16,58){\small (f)}
\end{overpic}
\caption{(Color online) (Color online) The pion FFs have been plotted with respect to $Q^2$ GeV$^2$ at different baryonic densities up to $\rho_B/\rho_0=4$ in Fig. (a) and (b). Both the in-medium and vacuum FFs have been compared with available vacuum experimental data \cite{NA7:1986vav,JeffersonLabFpi:2007vir,JeffersonLabFpi-2:2006ysh,JeffersonLabFpi:2000nlc,Dally:1982zk} in Fig. (c) along with lattice simulations \cite{QCDSFUKQCD:2006gmg,Ding:2024lfj} in Fig. (d) and (e). Fig. (f) shows the variation of the in-medium pion charge radius along with the squared charge radius with respect to the baryonic density $\rho_B/\rho_0$.}
\label{fig3}
\end{figure}

The EMFF gives information about the structure of the bound state system. The physical process that is associated with the EMFF of the pion is $\pi(P)+\gamma^{\ast}(\Delta)\rightarrow \pi(P^{\prime})$. The electromagnetic form factors can be obtained from the GPD $H$ by integrating over the longitudinal momentum fraction at zero skewness. For the pion, the chirally even GPD is given by,
\be
H(x,\xi,t)=\frac{1}{2}\int \frac{dz^{-}}{4\pi}e^{ixP^{+}z^{-}/2} \bra{\Pi(P^{\prime},\lambda^{\prime})}\bar{\vartheta}_{u}(0)\gamma^{+}\vartheta_{u}(z)\ket{\Pi(P,\lambda)}\vert_{z^{+}={\bf z}_{\perp}=0},
\ee
where momentum transfer $\Delta=P^{\prime}-P$ $(t\equiv\Delta^{2}\equiv -Q^{2})$ and skewness $\xi=-\frac{\Delta^{+}}{2P^{+}}$. We evaluate the form factor in terms of the GPD at $\xi=0$ as,
\begin{eqnarray}
 F(Q^{2})&=& e_{u} F^{u}(Q^{2})+e_{\bar{d}} F^{\bar{d}}(Q^{2})\nonumber\\
 &=&e_{u} \int H(x,0,t)~dx + e_{\bar{d}} \int H(1-x,0,t)~dx~.   
\end{eqnarray}
Fig.~\eqref{fig2}(a) represents the variation of the EMFFs $F(Q^{2})$ at different baryon densities upto $\rho_{B}/\rho_{0}=1$. The variation of the scaled EMFFs $Q^{2}F(Q^{2})$ for the same values of baryon densities has also been shown in Fig.~\eqref{fig2}(b). This essentially magnifies the difference between different baryon density curves for $Q^{2}>1$ GeV$^{2}$. Next in Fig.~\eqref{fig2}(c), we compare our results with the available vacuum experimental data~\cite{NA7:1986vav,JeffersonLabFpi:2007vir,JeffersonLabFpi-2:2006ysh,JeffersonLabFpi:2000nlc,Dally:1982zk} and with lattice calculations~\cite{QCDSFUKQCD:2006gmg,Ding:2024lfj} in Fig.~\eqref{fig2}(d) and (e). The same is shown in Fig.~\eqref{fig3}, albeit with larger values of baryon density (up to $4\rho_{0}$). We see a consistent suppression of in-medium EMFFs (across the momentum scale $Q^{2}\approx0-10$ GeV$^{2}$) as compared to vacuum EMFF derived within our hybrid LCQM-NJL approach. To end the discussion on the EMFFs, we show the important quantity charge radii in Fig.~\eqref{fig3}(f), which is defined with the help of EMFF as
\begin{eqnarray}
     \langle r_{\pi}^{2}\rangle&=&e_{u} \langle r_{u}^{2}\rangle + e_{\bar{d}} \langle r_{\bar{d}}^{2}\rangle\nonumber\\
     &=&- e_{u}~6\frac{\partial F^{u}(Q^{2})}{\partial Q^{2}}\vert_{Q^{2}\rightarrow 0}-e_{\bar{d}}~6\frac{\partial F^{\bar{d}}(Q^{2})}{\partial Q^{2}}\vert_{Q^{2}\rightarrow 0}~. 
\end{eqnarray}
We observe that the charge radii increase rapidly with increasing baryon density and slowly saturate around 0.6 fm as $\rho_{B}>3\rho_{0}$. 

\subsection{Parton Distribution Function (PDF)}
The PDF of the pion is the probability of finding the quark at a certain longitudinal momentum. It appears in the scattering cross-section of deep inclusive processes. The leading twist PDF for the pion can be obtained from the transverse momentum integrated leading twist TMD. The leading twist TMD for the quark in the pion is defined by,
\be
\Phi(x,{\bf k_{\perp}})=\frac{1}{2}\int \frac{d{z^{-}d^{2}{\bf z}_{\perp}}}{2(2\pi)^{3}} e^{ik\cdot z} \bra{\Pi(P,\lambda)}\bar{\vartheta}_{u}(0)\gamma^{+}\vartheta_{u}(z)\ket{\Pi(P,\lambda)}.
\ee
The unpolarized PDF obtained from the above TMD is given by,
\begin{eqnarray}
  f_{1}(x)&=&\int d^{2}{\bf k_{\perp}} \Phi(x,{\bf k_{\perp}})\nonumber\\
  &=&\int \frac{d^{2}{\bf k_{\perp}}}{16\pi^{3}}\left(|\psi(x,\bfk,\uparrow,\uparrow)|^{2}+|\psi(x,\bfk,\uparrow,\downarrow)|^{2}+|\psi(x,\bfk,\downarrow,\downarrow)|^{2}+|\psi(x,\bfk,\downarrow,\uparrow)|^{2}\right).\nonumber\\
\end{eqnarray}
\begin{figure*}[t]
\centering
\begin{overpic}[width=0.48\textwidth]{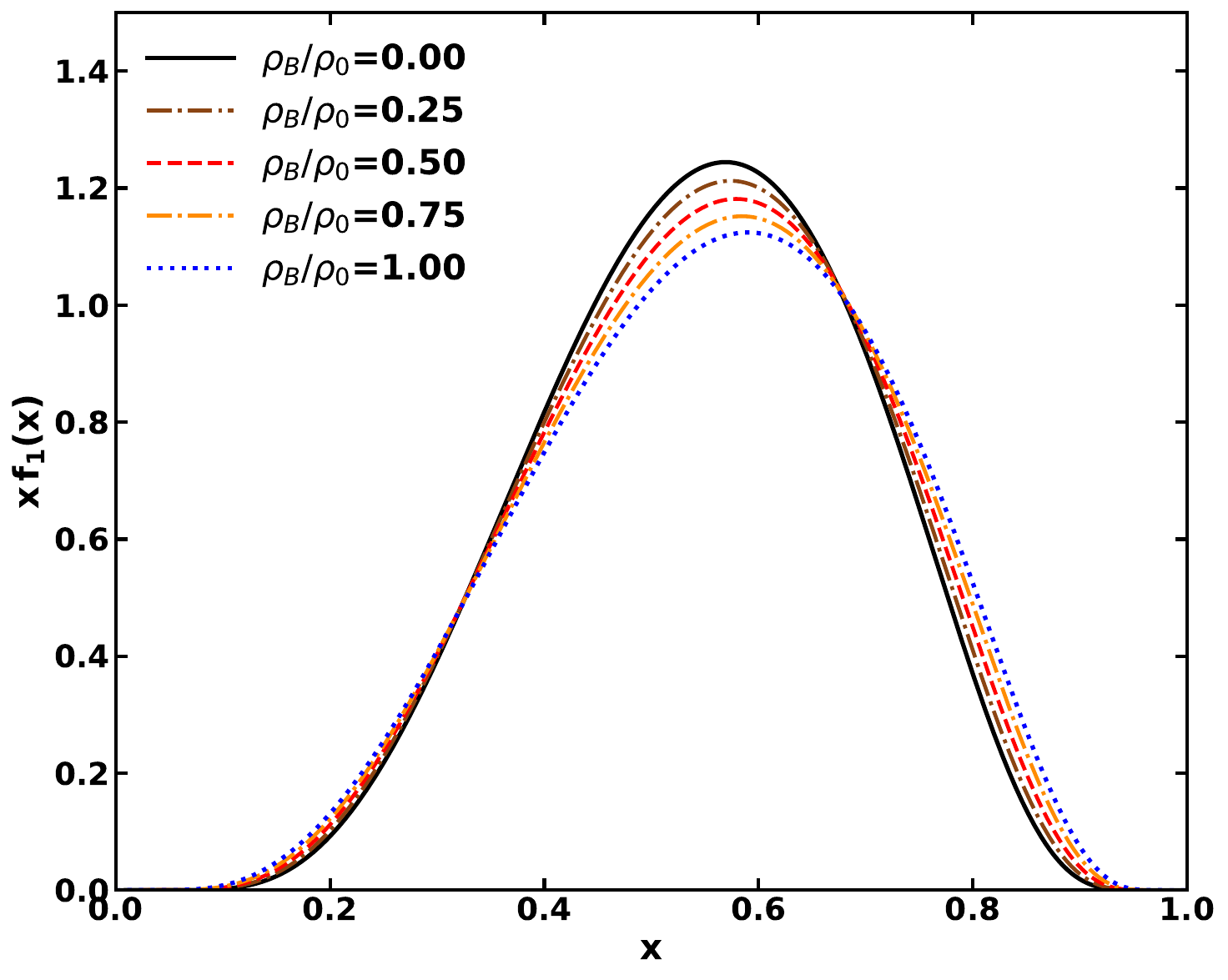}
    \put(85,58){\small (a)}
\end{overpic}
\hfill
\begin{overpic}[width=0.48\textwidth]{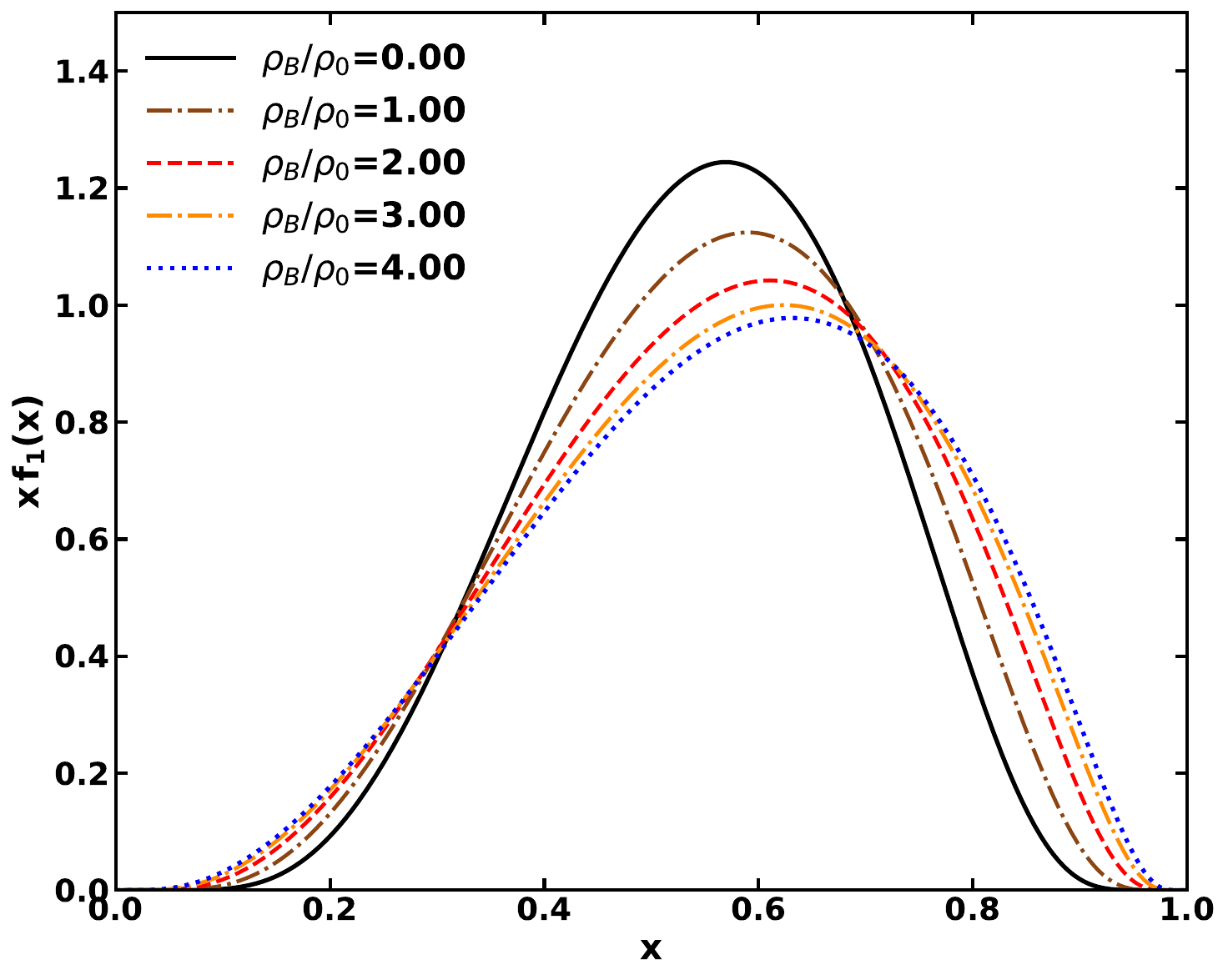}
    \put(85,58){\small (b)}
\end{overpic}

\begin{overpic}[width=0.48\textwidth]{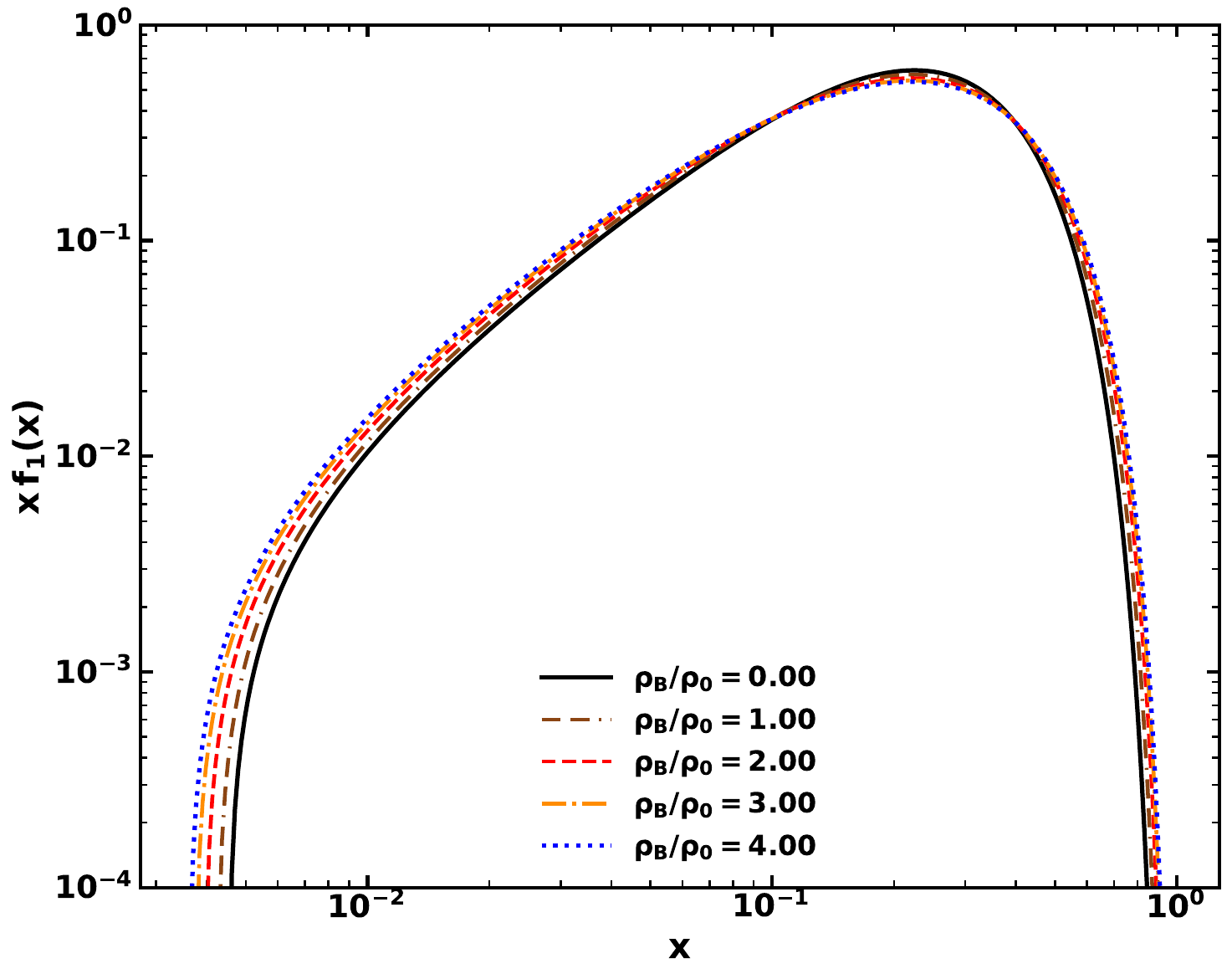}
    \put(16,58){\small (c)}
\end{overpic}
\caption{(Color online) The unpolarized pion PDFs $f_1(x)$ have been plotted at different baryonic densities with respect to $x$ in Fig. (a) and (b) at the model scale. In Fig. (c), we have plotted the in-medium evolved PDFs at $25$ GeV$^2$ using the NLO DGLAP equation.}
\label{fig4}
\end{figure*}
\begin{figure*}[t]
\centering
\begin{overpic}[width=0.48\textwidth]{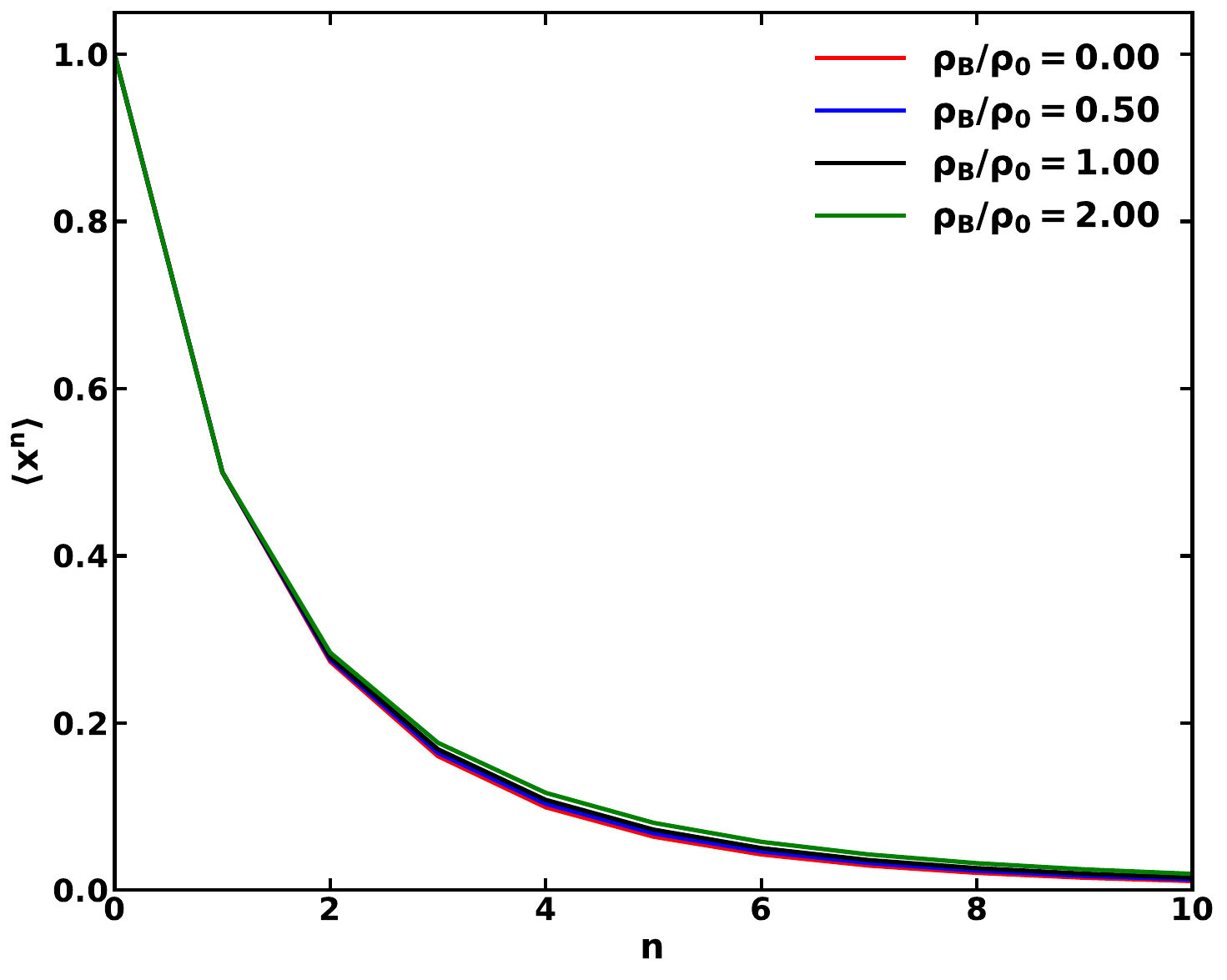}
    \put(17,58){\small (a)}
\end{overpic}
\hfill
\begin{overpic}[width=0.48\textwidth]{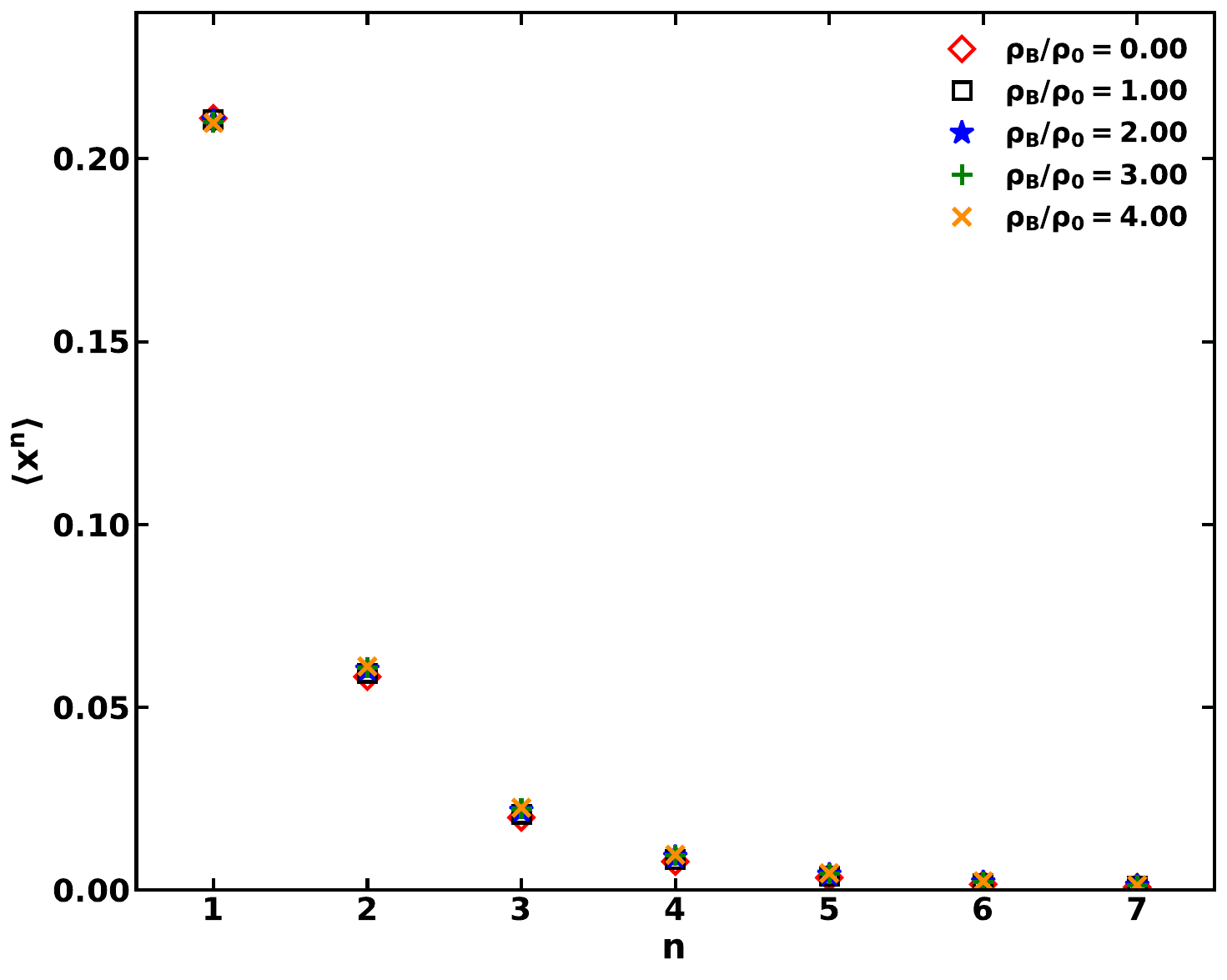}
    \put(17,56){\small (b)}
\end{overpic}
\caption{(Color online) The in-medium Mellin moment $\langle x^n \rangle$ at different baryonic densities has been calculated up to $n=10$ at the model scale in Fig. (a) and at $25$ GeV$^2$ in Fig. (b).}
\label{fig5}
\end{figure*}
In Fig.~\eqref{fig4}(a), we show the variation of the scaled PDF $xf_{1}(x)$ for quarks with respect to longitudinal momentum fraction $x$ at different baryon densities at the model scale. Fig.~\eqref{fig4}(b) contains the same information at higher baryon densities $\rho_{B}\leq 4\rho_{0}$. We also provide the PDF at zero baryon density (black curve) for the sake of comparison. From the quark PDF $f_{1}$ we can calculate the anti quark PDF $\bar{f}_{1}$ by replacing $x$ by $1-x$, i.e., $\bar{f}_{1}(x)=f(1-x)$.  The PDFs at model scale obey the sum rule $\int f_{1}(x)~dx=1$ and $\int \bar{f}_{1}(1-x)~dx=1$ . One observes that with increasing baryon density, the peak of the quark PDF shifts to the right, i.e., to higher values of $x$. This suggests an increase in the probability of finding the $u$-quark at higher longitudinal momentum inside a baryon-rich medium compared to the vacuum. Till now, the results of the PDF are shown at a model scale of 0.20 GeV$^{2}$. To see the longitudinal momentum fractions carried by quarks at a higher momentum scale, we evolved the u-quark PDF from model scale to a scale of $25$ GeV$^{2}$ using NLO DGLAP equations. The variation of evolved u-quark PDF has been shown in Fig.~\eqref{fig4}(c). See that the magnitude of the evolved PDF is smaller for most values of $x$ compared to the PDF evaluated at the model scale. In contrast to the PDF observed at the model scale, here we observe a very slight shift of the peak toward higher longitudinal momentum with increasing baryon density. Also, a very small variation among the PDFs at different baryon densities and vacuum is observed. From this behavior, one can conclude that the effect of the dense medium on the PDF is weaker at the high-momentum scale (perturbative regime) than at the non-perturbative scale. Subsequently, we display the in-medium Mellin moments $\langle x^{n}\rangle$ of quark PDFs, which encode non-perturbative aspects of QCD. The first moment ($n=1$) corresponds to the average momentum fraction
carried by the quark inside the pion. The higher-order Mellin moments give information about the quark densities with respect to different momentum fractions. The Mellin moments $\langle x^{n}\rangle$ are calculated as~\cite{Lu:2023yna},
\begin{equation}
    \langle x^{n}\rangle=\int x^{n}f_{1}(x) dx~.
\end{equation}
We see marginal changes in the Mellin moments with a change in baryon densities both in the model scale (Fig.~\eqref{fig5}(a)) and evolved scale (Fig.~\eqref{fig5}(b)). Also comparing with lattice QCD predictions of the pion Mellin moment at $Q=2$ GeV, we found that $\langle x \rangle=0.25$ compared to $\langle x \rangle=0.261$  \cite{Alexandrou:2020gxs}. We have also predicted the total valence quark contribution $2\langle x\rangle$ to the total momentum fraction of the pion in Table \ref{tab2} at different energy scales in vacuum. We found that at $Q^2=49$ GeV$^2$, only $40\%$ of the momentum fraction carried by the quark and antiquark, the remaining $60\%$ will be carried by the gluon and sea-quarks. At different energy scales, we have compared our results with JAM extraction \cite{Barry:2018ort}, xFitter \cite{Novikov:2020snp}, Lattice data \cite{Abdel-Rehim:2015owa,Best:1997qp,Detmold:2003tm,Martinelli:1987bh}, MAP \cite{Pasquini:2023aaf} and theoretical models \cite{Sutton:1991ay,Han:2018wsw,Gluck:1999xe,Ding:2019lwe,Wijesooriya:2005ir,Nam:2012vm,Watanabe:2017pvl,Lan:2019rba}. 
\begin{table}
    \centering
    \begin{tabular}{|c|c|c|c|c|c|c|}
    \hline
    & \multicolumn{6}{c|}{$Q^2$ (GeV$^2$)}  \\ \cline{2-7}
   $2\langle x \rangle$ & 1.69 & 4 &5 &5.76 & 27 &49 \\ \hline
    This work &  0.53 & 0.50 & 0.48 & 0.48 &0.42 & 0.40 \\ \hline
    JAM global fit \cite{Barry:2018ort} & 0.54 $\pm$ 0.01 & -& 0.48 $\pm$ 0.01& -& -& -\\ \hline
    JAM DY \cite{Barry:2018ort} & 0.60 $\pm$ 0.01 & & & & & \\ \hline
    xFitter (2020) \cite{Novikov:2020snp} & 0.55 $\pm$ 0.06 & 0.50 $\pm$ 0.05&0.49 $\pm$ 0.05 & 0.48 $\pm$ 0.05&0.42 $\pm$ 0.04& 0.41 $\pm$ 0.04\\ \hline
     MAP (2023) \cite{Pasquini:2023aaf} & 0.58 $\pm$ 0.03 & 0.52 $\pm$ 0.03 & 0.51 $\pm$ 0.03 & - & 0.45 $\pm$ 0.02 & - \\ \hline
    Latiice-1 \cite{Abdel-Rehim:2015owa} & - & 0.428 $\pm$ 0.03& - &- & -& -\\ \hline
    SMRS \cite{Sutton:1991ay}& - & 0.47& -& -& -& 0.49 $\pm$ 0.02\\ \hline
    Han \cite{Han:2018wsw}& - & 0.51 $\pm$ 0.03 &- &- & -& -\\ \hline
    GRVPI1 \cite{Gluck:1999xe} & - & 0.39&- &- &- &- \\ \hline
    Ding \cite{Ding:2019lwe} & - & 0.48 $\pm$0.03& -&- & -& -\\ \hline
    Lattice-2 \cite{Best:1997qp} & - &-& -& 0.558 $\pm$ 0.166 & -& -\\ \hline
      Lattice-3 \cite{Detmold:2003tm}& - & -& -& 0.48 $\pm$ 0.04& -& -\\ \hline
      WRH \cite{Wijesooriya:2005ir} &-  & -& -& -& 0.434 $\pm$0.022& -\\ \hline
       ChQM-1 \cite{Nam:2012vm} & - & -& -&- & 0.428& \\ \hline
        ChQM-2 \cite{Watanabe:2017pvl} & - & -& -&- & 0.46 & -\\ \hline
        Lattice-4 \cite{Martinelli:1987bh} & - & -& -&- & - & 0.46 $\pm$ 0.05\\ \hline
        BLFQ-NJL \cite{Lan:2019rba} & 0.54 $\pm$ 0.02 & 0.49 $\pm$ 0.018&- & 0.47 $\pm$ 0.018 & 0.42 $\pm$ 0.016 & 0.40 $\pm$ 0.015\\ \hline
    
    \end{tabular}
\caption{We have compared our total average longitudinal momentum fraction $2\langle x \rangle$ carried by the valence quark-antiquark at $Q^2=1.69$, $4$, $5$, $5.76$, $27$ and $49$ GeV$^2$ through NLO DGLAP evolution with JAM extraction \cite{Barry:2018ort}, xFitter \cite{Novikov:2020snp}, MAP23 \cite{,Pasquini:2023aaf}, Lattice data \cite{Abdel-Rehim:2015owa,Best:1997qp,Detmold:2003tm,Martinelli:1987bh} and theoretical models \cite{Sutton:1991ay,Han:2018wsw,Gluck:1999xe,Ding:2019lwe,Wijesooriya:2005ir,Nam:2012vm,Watanabe:2017pvl,Lan:2019rba}.}
    \label{tab2}
\end{table}

\section{Summary}\label{Sec:conclusion}
 To summarize the work, we have studied the in-medium properties of the pion by determining the quark distribution amplitude, electromagnetic form factor, and parton distribution function. To encode the dense-medium effects on the pion, we use the Nambu--Jona-Lasinio model in the symmetric nuclear medium and obtain the constituent quark mass as a function of baryon density. Whereas the quark-antiquark bound state (pion) wavefunction has been described by the light-cone quark model. We restrict ourselves to the valence Fock sector and express the light cone wave functions that implicitly depend on the baryon density of the surrounding medium via the constituent quark mass. We observe that the constituent quark mass decreases with an increase in baryon density as a result of partial restoration of chiral symmetry. Subsequently, we have solved the quark field correlators to get the distribution amplitude, electromagnetic form factor, and parton distribution function in terms of the light cone wave functions. In the presence of a dense medium, the distribution amplitude is observed to be suppressed for the intermediate longitudinal momentum fraction ($\sim$ 0.3-0.7) and enhanced at low or high longitudinal momentum fraction. The decay constant is observed to be less than the vacuum decay constant, consistent with the earlier predictions. The electromagnetic form factor in vacuum is observed to decrease with the momentum transfer, in par with several experimental and lattice studies. The in-medium form factor is suppressed from the corresponding vacuum values for the intermediate range of momentum transfer (0.1-1.5 GeV$^{2}$) and merges with the vacuum results for very low or high momentum transfer. In the limit of vanishing momentum transfer, the variation in the form factor is encoded in the charge radii of the pion. For relatively lower values of baryon densities ($\rho_{B}/\rho_{0}\sim 0-2$), we observe a rapid increase in the charge radii as a function of baryon density, and it slowly saturates at high baryon density. We have also analyzed the unpolarized parton distribution functions for pions in vacuum and in a medium with finite baryon density, both at the model scale $0.2$ GeV$^{2}$ and at the evolved scale of 25 GeV$^{2}$. To get the evolved parton distribution function, the next to leading order Dokshitzer–Gribov–Lipatov–Altarelli–Parisi equations are used. In a vacuum, the parton distribution function increases with the increase in longitudinal momentum fraction, has a peak (approximately around 1.2), and then decreases. It is observed that in a dense medium, for intermediate values of the longitudinal momentum fractions (approximately, 0.4-0.6), the parton distribution function is suppressed, whereas for other values it is enhanced.  The peak of the parton distribution function is also observed to shift towards higher longitudinal momentum fractions with an increase in the density of the surrounding medium. Although some similar medium effects persist in the evolved parton distribution functions, it is marginal compared to the model scale. The effect of the baryon density of the medium on the Mellin moments is seen to be minimal both in the model and perturbative scale. 

\section{Acknowledgement}
This work was partially supported by the Ministry of
Education (MoE), Govt. of India (A.D., S.P.); Board of Research in Nuclear Sciences (BRNS) and Department of Atomic Energy (DAE), Govt. of India, under Grant No. 57/14/01/2024-BRNS/313 (S.G.); and the Science and Engineering Research Board, Anusandhan-National Research Foundation, Government of India, under the scheme SERB-POWER Fellowship (Ref No. SPF/2023/000116) (H.D.).
\appendix
\section{Diquark and nucleon polarization}\label{ape1}
We first recognize from the Lagrangian~\eqref{AD2} that in the presence of vector fields, the kinetic momentum of the quarks is given by $k_{Q}^{\mu}\equiv k^{\mu}-2G_{\omega}\omega^{\mu}$. Therefore, one needs to replace $k^{\mu}$ by $k_{Q}^{\mu}$ and $(q-k)^{\mu}$ by $(q-k)^{\mu}_{Q}$ in $qq$ bubble graph expression as a result the propagators occuring in $\Pi_{s}(q)$ now change to $S(k)\rightarrow S(k_{Q})$ and $S(-(q-k))\rightarrow S(-(q-k)_{Q})=S(k_{Q}-q_{D})$ where $q_{D}^{\mu}\equiv q^{\mu}-4G_{\omega}\omega^{\mu}$. Similarly, in the quark-diquark bubble graph one makes the following changes: $S(k)\rightarrow S(k_{Q})$ and $\tau_{s}(q-k)\rightarrow  \tau_{s}((q-k)_{D})= \tau_{s}(q_{N}-k_{Q})$ where $q_{N}^{\mu}\equiv q^{\mu}-6G_{\omega}\omega^{\mu}$. To get the diquark mass, we rewrite Eq.~\eqref{AD4} in the presence of vector fields as
\be
\Pi_{s}(q_{D})&=& 6i \int \frac{d^{4}k_{Q}}{(2\pi)^{4}} \tr_{\rm D} [\gamma_{5}S(k_{Q})\gamma_{5}S(k_{Q}-q_{D})]\nonumber\\
&=& 6i \int \frac{d^{4}k_{Q}}{(2\pi)^{4}} \tr_{\rm D} \frac{\gamma_{5}(\slashed{k}_{Q}+m^{\ast})\gamma_{5}(\slashed{k}_{Q}-\slashed{q}_{D}+m^{\ast})}{(k_{Q}^{2}-m^{\ast 2})((k_{Q}-q_{D})^{2}-m^{\ast 2})}\nonumber\\
&=& 6i \int \frac{d^{4}k_{Q}}{(2\pi)^{4}}  \frac{-4k_{Q}^{2}+4k_{Q}\cdot q_{D}+4 m^{\ast 2}}{(k_{Q}^{2}-m^{\ast 2})((k_{Q}-q_{D})^{2}-m^{\ast 2})}~.\label{Ape1}
\ee
To regularize the diquark polarization integral, we employ the proper time regularization technique prescribed in Ref.~\cite{Bentz:2001vc}. For notational simplicity, we drop the subscripts from the diquark and quark momenta and transform them to the Euclidean space, i.e., we define the Euclidean momenta $p_{E}=(p^{0}_{E},\vec{p}_{E})$ for any momenta $p =(p^{0}=ip^{0}_{E},\vec{p}=\vec{p}_{E})$. Rewriting Eq.~\eqref{Ape1} with the transformed momenta we have,
\be
\Pi_{s}(q_{E})&=& -24 \int \frac{d^{4}k_{E}}{(2\pi)^{4}} \frac{k_{E}^{2}-k_{E}\cdot q_{E}+ m^{\ast 2}}{(k_{E}^{2}+m^{\ast 2})((k_{E}-q_{E})^{2}+m^{\ast 2})}\nonumber\\
&=&-24  \int \frac{d^{4}k_{E}^{\prime}}{(2\pi)^{4}} \frac{k_{E}^{\prime 2}-\frac{q_{E}^{2}}{4}+ m^{\ast 2}}{((k_{E}^{\prime} + \frac{q_{E}}{2})^{2}+m^{\ast 2})((k_{E}^{\prime} - \frac{q_{E}}{2})^{2}+m^{\ast 2})} ~(\text { where } k_{E}^{\prime}=k_{E}-\frac{q_{E}}{2})\nonumber\\
&=&-24  \int \frac{d^{4}k_{E}^{\prime}}{(2\pi)^{4}} \frac{1}{2} \bigg[ \frac{1}{(k_{E}^{\prime} + \frac{q_{E}}{2})^{2}+m^{\ast 2}}+ \frac{1}{(k_{E}^{\prime} - \frac{q_{E}}{2})^{2}+m^{\ast 2}}\bigg]\nonumber\\
&+& 12~q_{E}^{2}  \int \frac{d^{4}k_{E}^{\prime}}{(2\pi)^{4}} \frac{1}{((k_{E}^{\prime} + \frac{q_{E}}{2})^{2}+m^{\ast 2})((k_{E}^{\prime} - \frac{q_{E}}{2})^{2}+m^{\ast 2})}~.\label{Ape2}
\ee
To evaluate these integrals, one needs to use the following results~\cite{Bentz:2001vc},
\be
&&\int_{0}^{1} \frac{dx}{(xA+(1-x)B)^{2}} = \frac{1}{AB}~, \label{Ape3}\\
&& \frac{1}{A^{n}} \rightarrow \frac{1}{(n-1)!}\int_{1/\Lambda_{\rm UV}^{2}}^{1/\Lambda_{\rm IR}^{2}} d\tau~ \tau^{n-1} e^{-\tau A} ~(n\geq 1)~,\label{Ape4}\\
&& \ln A \rightarrow -\int_{1/\Lambda_{\rm UV}^{2}}^{1/\Lambda_{\rm IR}^{2}} \frac{d\tau}{\tau} e^{-\tau A}~.\label{Apextra}
\ee
One sees that Eq.~\eqref{Ape4} becomes exact in the limit $\Lambda_{\rm UV}\rightarrow \infty$ and $\Lambda_{\rm IR}\rightarrow 0$, and that these limits act as regularization. The first integral in Eq.~\eqref{Ape2} is performed by shifting the momentum variable and using Eq.~\eqref{Ape4}. Similarly, the second integral in Eq.~\eqref{Ape2} can be simplified by first using Eq.~\eqref{Ape3} and then Eq.~\eqref{Ape4}. The diquark polarization becomes,
\be
\Pi_{s}(q_{E}) &=&-24  \int \frac{d^{4}k_{E}^{\prime}}{(2\pi)^{4}}  \frac{1}{k_{E}^{\prime 2}+m^{\ast 2}}+ 12~q_{E}^{2}  \int \frac{d^{4}k_{E}^{\prime}}{(2\pi)^{4}} \int_{0}^{1}\frac{dx}{[x((k_{E}^{\prime} + \frac{q_{E}}{2})^{2}+m^{\ast 2})+(1-x)((k_{E}^{\prime} - \frac{q_{E}}{2})^{2}+m^{\ast 2})]^{2}}~\nonumber\\
&=&-\frac{3}{\pi^{2}}  \int dk_{E}^{\prime}~ k_{E}^{\prime 3}  \frac{1}{k_{E}^{\prime 2}+m^{\ast 2}} + 12~q_{E}^{2}  \int \frac{d^{4}k_{E}^{\prime}}{(2\pi)^{4}} \int_{0}^{1}\frac{dx}{[(k_{E}^{\prime}+(x-\frac{1}{2})q_{E})^{2}+x(1-x)q_{E}^{2}+m^{\ast 2}]^{2}}\nonumber\\
&=& -\frac{3}{\pi^{2}} \int dk_{E}^{\prime}~ k_{E}^{\prime 3} \int_{1/\Lambda_{\rm UV}^{2}}^{1/\Lambda_{\rm IR}^{2}} e^{-\tau(k_{E}^{\prime 2}+m^{\ast 2})} d\tau + 12~q_{E}^{2}  \int \frac{d^{4}k_{E}^{\prime}}{(2\pi)^{4}} \int_{0}^{1}\frac{dx}{[k_{E}^{\prime 2}+x(1-x)q_{E}^{2}+m^{\ast 2}]^{2}}\nonumber\\
&&(\text{where we shift the variable in the second integral } k_{E}^{\prime}+(x-\frac{1}{2})q_{E} \rightarrow k_{E}^{\prime})\nonumber\\
&=& -\frac{3}{\pi^{2}} \int dk_{E}^{\prime}~ k_{E}^{\prime 3} \int_{1/\Lambda_{\rm UV}^{2}}^{1/\Lambda_{\rm IR}^{2}} e^{-\tau(k_{E}^{\prime 2}+m^{\ast 2})}~ d\tau + \frac{3q_{E}^{2}}{2\pi^{2}}  \int dk_{E}^{\prime}~ k_{E}^{\prime 3} \int_{0}^{1} dx \int_{1/\Lambda_{\rm UV}^{2}}^{1/\Lambda_{\rm IR}^{2}} \tau~ d\tau e^{-\tau (k_{E}^{\prime 2}+x(1-x)q_{E}^{2}+m^{\ast 2})}\nonumber\\
&=& -\frac{3}{2\pi^{2}}  \int_{1/\Lambda_{\rm UV}^{2}}^{1/\Lambda_{\rm IR}^{2}} \frac{d\tau}{\tau^{2}} e^{-\tau m^{\ast 2}} + \frac{3q_{E}^{2}}{4\pi^{2}} \int_{0}^{1} dx \int_{1/\Lambda_{\rm UV}^{2}}^{1/\Lambda_{\rm IR}^{2}} \frac{d\tau}{\tau} e^{-\tau (x(1-x)q_{E}^{2}+m^{\ast 2})}\nonumber\\
&=& -\frac{3}{2\pi^{2}}  \int_{1/\Lambda_{\rm UV}^{2}}^{1/\Lambda_{\rm IR}^{2}} \frac{d\tau}{\tau^{2}} e^{-\tau m^{\ast 2}}+ \frac{3q_{E}^{2}}{4\pi^{2}} \int_{1/\Lambda_{\rm UV}^{2}}^{1/\Lambda_{\rm IR}^{2}} \frac{d\tau}{\tau}  \bigg[ xe^{-\tau (x(1-x)q_{E}^{2}+m^{\ast 2})}\Big|_{0}^{1}+\int_{0}^{1} dx ~ x(1-2x)\tau q_{E}^{2} e^{-\tau (x(1-x)q_{E}^{2}+m^{\ast 2})}\bigg]\nonumber\\
&=& -\frac{3}{2\pi^{2}}  \int_{1/\Lambda_{\rm UV}^{2}}^{1/\Lambda_{\rm IR}^{2}} \frac{d\tau}{\tau^{2}} e^{-\tau m^{\ast 2}} + \frac{3q_{E}^{2}}{4\pi^{2}} \int_{1/\Lambda_{\rm UV}^{2}}^{1/\Lambda_{\rm IR}^{2}} \frac{d\tau}{\tau}  e^{-\tau m^{\ast 2}} +  \frac{3q_{E}^{4}}{4\pi^{2}} \int_{0}^{1} dx ~ x(1-2x)\int_{1/\Lambda_{\rm UV}^{2}}^{1/\Lambda_{\rm IR}^{2}} d\tau  e^{-\tau (x(1-x)q_{E}^{2}+m^{\ast 2})}~.\nonumber\\
\label{Ape5}
\ee
We can easily express the above equation as
\be
&&\Pi_{s}(q_{E}) = -\frac{3}{4\pi^{2}}\left(2C_{2}(m^{\ast 2})-q_{E}^{2}C_{1}(m^{\ast 2})+q_{E}^{4} \int_{0}^{1} dx~ x(1-2x) \frac{e^{-A_{D}/\Lambda_{\rm IR}^{2}}-e^{-A_{D}/\Lambda_{\rm UV}^{2}}}{A_{D}} \right)~,\label{Ape6}
\ee
where we define $C_{n}(m^{\ast 2})\equiv \int_{1/\Lambda_{\rm UV}^{2}}^{1/\Lambda_{\rm IR}^{2}} \frac{d\tau}{\tau^{n}} e^{-\tau m^{\ast 2}}$ and $A_{D}(m^{\ast},q_{E})\equiv m^{\ast 2}+x(1-x)q_{E}^{2}$. Transforming back to the Minkowski space from the Euclidean space, we have,
\be
&&\Pi_{s}(q_{D}^{2}) = -\frac{3}{4\pi^{2}}\left(2C_{2}(m^{\ast 2})+q_{D}^{2}C_{1}(m^{\ast 2})+q_{D}^{4} \int_{0}^{1} dx~ x(1-2x) \frac{e^{-A_{D}/\Lambda_{\rm IR}^{2}}-e^{-A_{D}/\Lambda_{\rm UV}^{2}}}{A_{D}} \right)~,\label{Ape7}
\ee
where $A_{D}(m^{\ast},q_{D})= m^{\ast 2}-x(1-x)q_{D}^{2}$, and we retain the diquark label in the momentum variable. We note that Eq.~\eqref{Ape7} is identical to Eq.~(B2) of Ref.~\cite{Bentz:2001vc}, except for the power of the diquark momentum appearing in the third term. In Ref.~\cite{Bentz:2001vc}, this power was given as two; however, this appears to be a typographical error, as can be readily verified through dimensional analysis. The mass of the diquark $m_{D}=m_{D}(m^{\ast})$ is calculated with the help of Eq.~\eqref{Ape7} using  $1+2G_{s}\Pi_{s}(q_{D}^{2}=m_{D}^{2})=0$.

To evaluate the quark-diquark bubble graph or nucleonic polarization, we follow Refs.~\cite{Bentz:2001vc,Mineo:2003vc} and approximate the diquark $T$-matrix $\tau_{s}(k_{D})=4iG_{s}-\frac{ig_{s}}{k_{D}^{2}-m_{D}^{2}}$, where $g_{s}^{-1}=-\frac{1}{2}\frac{d \Pi_{s}(q^{2}_{D})}{d q^{2}_{D}}\Big|_{q_{D}^{2}=m_{D}^{2}}$. The polarization for the nucleon is given by,
\be
\Pi_{N}(q_{N})&=&- \int \frac{d^{4}k_{Q}}{(2\pi)^{4}} [S(k_{Q})\tau_{s}(q_{N}-k_{Q})]\nonumber\\
&=&- \int \frac{d^{4}k_{Q}}{(2\pi)^{4}} \frac{1}{\slashed{k}_{Q}-m^{\ast}} \left(4iG_{s}-\frac{ig_{s}}{(q_{N}-k_{Q})^{2}-m_{D}^{2}}\right)\nonumber\\
&=& -\int \frac{d^{4}k_{Q}}{(2\pi)^{4}}  \frac{4iG_{s}(\slashed{k}_{Q}+m^{\ast})}{(k_{Q}^{2}-m^{\ast 2})}+\int \frac{d^{4}k_{Q}}{(2\pi)^{4}}  \frac{ig_{s}(\slashed{k}_{Q}+m^{\ast})}{(k_{Q}^{2}-m^{\ast 2})((q_{N}-k_{Q})^{2}-m_{D}^{2})}\nonumber\\
&=&-\int \frac{d^{4}k_{Q}}{(2\pi)^{4}}  \frac{4iG_{s}m^{\ast}}{(k_{Q}^{2}-m^{\ast 2})}+\int \frac{d^{4}k_{Q}}{(2\pi)^{4}}  \frac{ig_{s}(\slashed{k}_{Q}+m^{\ast})}{(k_{Q}^{2}-m^{\ast 2})((q_{N}-k_{Q})^{2}-m_{D}^{2})}~\nonumber\\
&=&-\int \frac{d^{4}k_{Q}}{(2\pi)^{4}}  \frac{4iG_{s}m^{\ast}}{(k_{Q}^{2}-m^{\ast 2})}+\int \frac{d^{4}k_{Q}}{(2\pi)^{4}}  \frac{ig_{s}(\slashed{k}_{Q}+\frac{\slashed{q}_{N}}{2}+m^{\ast})}{(k_{Q}+\frac{q_{N}}{2})^{2}-m^{\ast 2})((k_{Q}-\frac{q_{N}}{2})^{2}-m_{D}^{2})}~,\label{Ape8}
\ee
where to obtain the last line, we shifted the variable in the second integral $k_{Q}\rightarrow k_{Q}+\frac{q_{N}}{2}$.
To simplify further, we Wick rotate the above integral to the Euclidean space (and drop the additional momentum subscripts for notional simplicity), the gamma matrices transform similarly to momentum, i.e., $\gamma^{0}=i\gamma^{0}_{E}$,
\be
\Pi_{N}(q_{E})&=&-\int \frac{d^{4}k_{E}}{(2\pi)^{4}}  \frac{4G_{s}m^{\ast}}{(k_{E}^{2}+m^{\ast 2})}+\int \frac{d^{4}k_{E}}{(2\pi)^{4}}  \frac{g_{s} (\slashed{k}_{E}+\frac{\slashed{q}_{E}}{2}-m^{\ast})}{(k_{E}+\frac{q_{E}}{2})^{2}+m^{\ast 2})((k_{E}-\frac{q_{E}}{2})^{2}+m_{D}^{2})}~,\nonumber\\
&=& -\int \frac{d^{4}k_{E}}{(2\pi)^{4}}  \frac{4G_{s}m^{\ast}}{(k_{E}^{2}+m^{\ast 2})}+\int \frac{d^{4}k_{E}}{(2\pi)^{4}} \int_{0}^{1} \frac{dx ~g_{s}(\slashed{k}_{E}+\frac{\slashed{q}_{E}}{2}-m^{\ast})}{[(1-x)(k_{E}+\frac{q_{E}}{2})^{2}+m^{\ast 2})+x(k_{E}-\frac{q_{E}}{2})^{2}+m_{D}^{2})]^{2}}\nonumber\\
&=& -\int \frac{d^{4}k_{E}}{(2\pi)^{4}}  \frac{4G_{s}m^{\ast}}{(k_{E}^{2}+m^{\ast 2})}+\int \frac{d^{4}k_{E}}{(2\pi)^{4}} \int_{0}^{1} \frac{dx ~g_{s}(\slashed{k}_{E}+\frac{\slashed{q}_{E}}{2}-m^{\ast})}{[(k_{E}-(x-\frac{1}{2})q_{E})^{2}+x(1-x)q_{E}^{2}+(1-x)m^{\ast 2}+x m_{D}^{2}]^{2}}\nonumber\\
&=&-\int \frac{d^{4}k_{E}}{(2\pi)^{4}}  \frac{4G_{s}m^{\ast}}{(k_{E}^{2}+m^{\ast 2})}+\int \frac{d^{4}k_{E}}{(2\pi)^{4}} \int_{0}^{1} \frac{dx ~g_{s}(\slashed{k}_{E}+x\slashed{q}_{E}-m^{\ast})}{[k_{E}^{2}+x(1-x)q_{E}^{2}+(1-x)m^{\ast 2}+x m_{D}^{2}]^{2}}~,\label{Ape9}
\ee
where to obtain the last line, we shift the integration variable in the second integral for a fixed $x$ as $k_{E}-(x-\frac{1}{2})q_{E} \rightarrow k_{E}$. Using Eq.~\eqref{Ape4} in Eq.~\eqref{Ape9} we have,
\be
\Pi_{N}(q_{E})&=& -\frac{G_{s}m^{\ast}}{4\pi^{2}}C_{2}(m^{\ast 2})+\frac{g_{s}}{8\pi^{2}}\int k_{E}^{3}dk_{E}\int \tau d\tau \int_{0}^{1} dx~(x\slashed{q}_{E}-m^{\ast})~e^{-\tau(k_{E}^{2}+x(1-x)q_{E}^{2}+m^{\ast 2}+x(m_{D}^{2}-m^{\ast 2}))}.\label{Ape10}
\ee
The second term in Eq.~\eqref{Ape10} can be simplified by performing an integration by parts over the variable $x$ by treating the exponential as the first function and the other term $x\slashed{q}_{E}-m^{\ast}$ as the second to give,  
\be
\Pi_{N}(q_{E})&=& -\frac{G_{s}m^{\ast}}{4\pi^{2}}C_{2}(m^{\ast 2})\nonumber\\
&+&\frac{g_{s}}{8\pi^{2}} \int \tau d\tau \int k_{E}^{3}dk_{E} \bigg[\left(\frac{\slashed{q}_{E}}{2}-m^{\ast}\right)e^{-\tau(k_{E}^{2}+m_{D}^{2})} +\tau\int ((1-2x)q_{E}^{2}+m_{D}^{2}-m^{\ast 2}) e^{-\tau(k_{E}^{2}+A_{N})}~x(\frac{x}{2}\slashed{q}_{E}-m^{\ast}) dx\bigg] .\nonumber\\
&=&-\frac{G_{s}m^{\ast}}{4\pi^{2}}C_{2}(m^{\ast 2})+\frac{g_{s}}{16\pi^{2}}\left(\frac{\slashed{q}_{E}}{2}-m^{\ast}\right)~C_{1}(m_{D}^{2})\nonumber\\
&-&\frac{g_{s}}{16\pi^{2}}\int_{0}^{1} dx~ (m_{D}^{2}-m^{\ast 2}+(1-2x)q_{E}^{2})~ x\left(\frac{x}{2}\slashed{q}_{E}-m^{\ast}\right)\left(\frac{e^{-A_{N}/\Lambda_{\rm IR}^{2}}-e^{-A_{N}/\Lambda_{\rm UV}^{2}}}{A_{N}} \right)~,
\label{Ape11}
\ee
where $A_{N}\equiv m^{\ast 2}+x(m_{D}^{2}-m^{\ast 2})+x(1-x)q_{E}^{2}$. Moving back to the Minkowski space and retaining the subscript for nucleon momentum, we have,
\be
\Pi_{N}(q_{N})&=&-\frac{G_{s}m^{\ast}}{4\pi^{2}}C_{2}(m^{\ast 2})-\frac{g_{s}}{16\pi^{2}}\left(\frac{\slashed{q}_{N}}{2}+m^{\ast}\right)~C_{1}(m_{D}^{2})\nonumber\\
&+&\frac{g_{s}}{16\pi^{2}}\int_{0}^{1} dx~ (m_{D}^{2}-m^{\ast 2}-(1-2x)q_{N}^{2})~ x\left(\frac{x}{2}\slashed{q}_{N}+m^{\ast}\right)\left(\frac{e^{-A_{N}/\Lambda_{\rm IR}^{2}}-e^{-A_{N}/\Lambda_{\rm UV}^{2}}}{A_{N}} \right)~.
\label{Ape12}
\ee
Eq.~\eqref{Ape12} matches exactly with Eq. (B3) of Ref.~\cite{Bentz:2001vc}.
The mass of the nucleon $m_{N}=m_{N}(m^{\ast})$ is calculated with the help of Eq.~\eqref{Ape12} using  $1+\frac{3}{m^{\ast}}\Pi_{N}(\slashed{q}_{N}=m_{N})=0$. It should be noted that the nucleon momenta in the presence of vector field is $q^{\mu}=q_{N}^{\mu}+6G_{\omega}\omega^{\mu}$ so that $\epsilon\equiv q^{0}=q^{0}_{N}+6G_{\omega}\omega^{0}$. Using $q_{N}^{2}=m_{N}^{2}$, the energy of the nucleons in the presence of the vector fields becomes $\epsilon=6G_{\omega}\omega^{0}+E_{N}$ where $E_{N}\equiv\sqrt{\vec{q}_{N}^{2}+m_{N}^{2}}$.

\section{Grand potential and the gap equations}\label{ape2}
We simplify the expression of the grand potential given in Eq.~\eqref{AD8} and give its explicit form. Using the Fermi distribution for the nucleons $f=1/[e^{(\epsilon-\mu_{B})/T}+1]\xrightarrow{T=0} \theta(\mu_{B}^{\ast}-E_{N})$ where $\mu_{B}^{\ast}=\mu_{B}-6G_{\omega}\omega^{0}$ we express the grand potential as sum three terms: vacuum corresponding to the quark loops and scalar mean field ($\Omega_{q}$), the term corresponding the vector field ($\Omega_{\rm vec}$) and nucleonic Fermi sea ($\Omega_{N}$),
\be
\Omega = \Omega_{q}+\Omega_{\rm vec}+\Omega_{N} \text{ where, }\Omega_{q}&=& i\gamma_{q}\int \frac{d^{4}k}{(2\pi)^{4}} \ln \frac{k^{2}-m^{\ast 2}+i\epsilon}{k^{2}-m_{0}^{\ast 2}+i\epsilon} + \frac{(m^{\ast}-m)^{2}}{4G_{\pi}}-\frac{(m^{\ast}_{0}-m)^{2}}{4G_{\pi}}\nonumber~,\\
\Omega_{\rm vec}&=&-G_{\omega}\omega_{0}^{2}\nonumber~,\\
\Omega_{N}&=& \gamma_{N}\int \frac{d^{3}\vec{k}_{N}}{(2\pi)^{3}}  \theta(\mu_{B}^{\ast}-E_{N}) E_{N} -\mu^{\ast}_{B} \rho_{B}~.\label{Ape13}
\ee
We simplify the quark and nucleon contributions, first writing the quark contribution and using Eq.~\eqref{Apextra},
\be
\Omega_{q} &=& i\gamma_{q}\int \frac{d^{4}k}{(2\pi)^{4}} \ln \frac{k^{2}-m^{\ast 2}+i\epsilon}{k^{2}-m_{0}^{\ast 2}+i\epsilon} + \frac{(m^{\ast}-m)^{2}}{4G_{\pi}}-\frac{(m^{\ast}_{0}-m)^{2}}{4G_{\pi}}\nonumber\\
&=& -\gamma_{q}\int \frac{d^{4}k_{E}}{(2\pi)^{4}} [\ln(k_{E}^{2}+m^{\ast 2})-\ln(k_{E}^{2}+m_{0}^{2})] + \frac{(m^{\ast}-m)^{2}}{4G_{\pi}}-\frac{(m^{\ast}_{0}-m)^{2}}{4G_{\pi}}\nonumber\\
&=& \frac{\gamma_{q}}{8\pi^{2}}\int_{1/\Lambda_{\rm UV}^{2}}^{1/\Lambda_{\rm IR}^{2}}  \frac{d\tau}{\tau}\int k_{E}^{3}dk_{E}~\bigg[e^{-\tau(k_{E}^{2}+m^{\ast 2})}-e^{-\tau(k_{E}^{2}+m^{2}_{0})}\bigg] + \frac{(m^{\ast}-m)^{2}}{4G_{\pi}}-\frac{(m^{\ast}_{0}-m)^{2}}{4G_{\pi}}\nonumber\\
&=& \frac{\gamma_{q}}{16\pi^{2}}\int_{1/\Lambda_{\rm UV}^{2}}^{1/\Lambda_{\rm IR}^{2}}  \frac{d\tau}{\tau^{3}}\bigg[e^{-\tau m^{\ast 2}}-e^{-\tau m^{2}_{0}}\bigg] + \frac{(m^{\ast}-m)^{2}}{4G_{\pi}}-\frac{(m^{\ast}_{0}-m)^{2}}{4G_{\pi}}\label{Ape14}~.
\ee
To evaluate the contribution from the nucleonic Fermi sea, we make the change of variables $k_{N}=m_{N}\sinh x$, so that $E_{N}=m_{N}\cosh x$ along the way,
\be
\Omega_{N}&=& \frac{\gamma_{N}}{2\pi^{2}}\int_{0}^{k_{N}^{F}} k_{N}^{2} E_{N}  dk_{N} -\mu^{\ast}_{B} \frac{\gamma_{N}}{2\pi^{2}}\int_{0}^{k_{N}^{F}} k_{N}^{2} dk_{N}, (k_{N}^{F}=\sqrt{\mu_{B}^{\ast 2}-m_{N}^{2}})\nonumber\\
&=& \frac{\gamma_{N}}{2\pi^{2}}\int m_{N}^{4} \sinh^{2}x\cosh^{2}x ~dx -\mu^{\ast}_{B} \frac{\gamma_{N}}{6\pi^{2}}~(k_{N}^{F})^{3}\mu_{B}^{\ast},\nonumber\\
&=& \frac{\gamma_{N}}{2\pi^{2}} \left (\frac{1}{4}k_{N}^{F}\mu_{B}^{\ast 3}-\frac{1}{8}m_{N}^{2}k_{N}^{F}\mu_{B}^{\ast}-\frac{1}{8}m_{N}^{4}\sinh^{-1} \frac{k_{N}^{F}}{m_{N}}-\frac{1}{3}(k_{N}^{F})^{3}\mu_{B}^{\ast}\right)\label{Ape15},
\ee
where we used the result $\int \sinh^{2}x\cosh^{2}x ~dx=\frac{1}{4}\sinh x\cosh^{3}x-\frac{1}{8}\sinh x\cosh x-\frac{x}{8}$. Now, one can directly arrive at the gap equations Eqs.~\eqref{AD11} and \eqref{AD12} by minimizing the grand potential (see Eqs.~\eqref{AD9} and \ref{AD10}),
\be
\left(\frac{\partial \Omega}{\partial \omega_{0}}\right)_{\mu_{B},m^{\ast}}&=&\left(\frac{\partial \Omega_{q}}{\partial \omega_{0}}\right)_{\mu_{B},m^{\ast}}+\left(\frac{\partial \Omega_{\rm vec}}{\partial \omega_{0}}\right)_{\mu_{B},m^{\ast}}+\left(\frac{\partial \Omega_{N}}{\partial \omega_{0}}\right)_{\mu_{B},m^{\ast}}
\nonumber\\
&=&-2G_{\omega}\omega_{0}+\frac{\partial}{\partial \omega_{0}}\frac{\gamma_{N}}{2\pi^{2}} \left (\frac{1}{4}k_{N}^{F}\mu_{B}^{\ast 3}-\frac{1}{8}m_{N}^{2}k_{N}^{F}\mu_{B}^{\ast}-\frac{1}{8}m_{N}^{4}\sinh^{-1} \frac{k_{N}^{F}}{m_{N}}-\frac{1}{3}(k_{N}^{F})^{3}\mu_{B}^{\ast}\right)\nonumber\\
&=&-2G_{\omega}\omega_{0}+2G_{\omega}\frac{\gamma_{N}}{2\pi^{2}}(k_{N}^{F})^{3}\label{Ape16}
\ee
where we used $\frac{\partial k_{N}^{F}}{\partial \omega_{0}}=-6G_{\omega}\frac{\mu_{B}^{\ast}}{k_{N}^{F}}$, $\frac{\partial \mu_{B}^{\ast}}{\partial \omega_{0}}=-6G_{\omega}$ and $\frac{\partial}{\partial \omega_{0}}\sinh^{-1} \frac{k_{N}^{F}}{m_{N}}=\frac{-6G_{\omega}}{k_{N}^{F}}$. One easily obtains Eq.~\eqref{AD11} from Eq.~\eqref{Ape16} by identifying $\rho_{B}=\frac{\gamma_{N}}{2\pi^{2}}\frac{(k_{N}^{F})^{3}}{3}$. Similarly, the gap equation Eq.~\eqref{AD12} is obtained by minimizing grand potential with respect to the constituent quark mass,
\be
\left(\frac{\partial \Omega}{\partial m^{\ast}}\right)_{\mu_{B},\omega_{0}}&=&\left(\frac{\partial \Omega_{q}}{\partial m^{\ast}}\right)_{\mu_{B},\omega_{0}}+\left(\frac{\partial \Omega_{\rm vec}}{\partial m^{\ast}}\right)_{\mu_{B},\omega_{0}}+\left(\frac{\partial \Omega_{N}}{\partial m^{\ast}}\right)_{\mu_{B},\omega_{0}}
\nonumber\\
&=&-\frac{m^{\ast}\gamma_{q}}{8\pi^{2}}\int_{\Lambda_{\rm UV}^{2}}^{\Lambda_{\rm IR}^{2}} \frac{d\tau}{\tau^{2}} e^{-\tau m^{\ast 2}} + \frac{m^{\ast} -m}{2G_{\pi}}+\frac{\partial}{\partial m^{\ast}}\frac{\gamma_{N}}{2\pi^{2}} \left (\frac{k_{N}^{F}\mu_{B}^{\ast 3}}{4}-\frac{m_{N}^{2}k_{N}^{F}\mu_{B}^{\ast}}{8}-\frac{m_{N}^{4}}{8}\sinh^{-1} \frac{k_{N}^{F}}{m_{N}}-\frac{(k_{N}^{F})^{3}\mu_{B}^{\ast}}{3}\right)\nonumber\\
&=&-\frac{m^{\ast}\gamma_{q}}{8\pi^{2}}\int_{\Lambda_{\rm UV}^{2}}^{\Lambda_{\rm IR}^{2}} \frac{d\tau}{\tau^{2}} e^{-\tau m^{\ast 2}} + \frac{m^{\ast} -m}{2G_{\pi}}+\frac{\gamma_{N}}{4\pi^{2}}\left(k_{N}^{F}\mu_{B}^{\ast}-m_{N}^{2}\sinh^{-1} \frac{k_{N}^{F}}{m_{N}}\right)m_{N}\frac{\partial m_{N}}{\partial m^{\ast}}~,\label{Ape17}
\ee
where we used $\frac{\partial k_{N}^{F}}{\partial m^{\ast}}=-\frac{m_{N}}{k_{N}^{F}}\frac{\partial m_{N}}{\partial m^{\ast}}$ and $\frac{\partial}{\partial m^{\ast}}\sinh^{-1} \frac{k_{N}^{F}}{m_{N}}=-\frac{\mu_{B}^{\ast}}{k_{N}^{F}m_{N}}\frac{\partial m_{N}}{\partial m^{\ast}}$. Equating Eq.~\eqref{Ape17} to zero, one easily obtains the gap equation~\eqref{AD12}.
\bibliography{ref_lc_revised}

\end{document}